\documentclass{article}

\usepackage{main}

\usepackage[utf8]{inputenc} 
\usepackage[T1]{fontenc}    
\usepackage{hyperref}       
\usepackage{url}            
\usepackage{booktabs}       
\usepackage{amsfonts}       
\usepackage{nicefrac}       
\usepackage{microtype}      
\usepackage{lipsum}		
\usepackage{graphicx}
\usepackage{natbib}
\usepackage{doi}
\usepackage{multirow}%
\usepackage{amsmath,amssymb}%
\usepackage{amsthm}%
\usepackage{mathrsfs}%
\usepackage[figuresright]{rotating}%
\usepackage[title]{appendix}%
\usepackage{xcolor}%
\usepackage{textcomp}%
\usepackage{manyfoot}%
\usepackage{booktabs}%
\usepackage{algorithm}%
\usepackage{algorithmicx}%
\usepackage{algpseudocode}%
\usepackage{program}%
\usepackage{listings}%

\title{Snapshot Averaged Matrix Pencil Method\,(SAM) For Direction of Arrival Estimation}

\author{ {Harsha Avinash Tanti$^*$, Abhirup Datta} \\
	Department of Astronomy, Astrophysics and Space Engineering\\
	Indian Institute of Technology Indore\\
	Indore, Madhya Pradesh, India 453552 \\
	\texttt{phd190112009@iiti.ac.in, abhirup.datta@iiti.ac.in} \\
	\And
	{S. Ananthakrishnan} \\
	Department of Electronic Science\\
	Savitribai Phule Pune University\\
	Pune, Maharashtra, India 411007 \\
	\texttt{subra.anan@gmail.com} \\
}

\date{}

\renewcommand{\shorttitle}{SAM based DoA Algorithm}

\begin{document}
\maketitle

\begin{abstract}
The estimation of the direction of electromagnetic\,(EM) waves from a radio source using electrically short antennas is one of the challenging problems in the field of radio astronomy. 
In this paper we have developed an algorithm which performs better in direction and polarization estimations than the existing algorithms. 
Our proposed algorithm Snapshot Averaged Matrix Pencil Method\,(SAM) is a modification to the existing Matrix Pencil Method\,(MPM) based Direction of Arrival\,(DoA) algorithm. 
In general, MPM estimates DoA of the incoherent EM waves in the spectra using unitary transformations and least square method\,(LSM). 
Our proposed SAM modification is made in context to the proposed Space Electric and Magnetic Sensor\,(SEAMS) mission to study the radio universe below 16\,MHz. 
SAM introduces a snapshot averaging method to improve the incoherent frequency estimation improving the accuracy of estimation. 
It can also detect polarization to differentiate between Right Hand Circular Polarlization\,(RHCP), Right Hand Elliptical Polarlization\,(RHEP), Left Hand Circular Polarlization\,(LHCP), Left Hand Elliptical Polarlization\,(LHEP) and Linear Polarlization\,(LP). 
This paper discusses the formalism of SAM and shows the initial results of a scaled version of a DoA experiment at a resonant frequency of $\sim$72\,MHz.
\end{abstract}


The radio frequencies ranging from 0.3 to 16\,MHz is one of the unexplored realms of the electromagnetic spectrum in the field of radio astronomy. 
This frequency range covers the red-shifted 21\,cm line from the early Universe ($\sim$ 0.38 to 400 million years after the Big Bang), radio emissions from planetary and exoplanetary magnetosphere, and traces a wide range of astrophysical phenomena \citep{Bentum2018ieee, Zarka2007, Bentum2017ieee, Rajan2016, Bentum2020}. 
In this frequency range the ground-based astronomical observations have been infrequent, due to the presence of the Earth's ionosphere and radio frequency interference (RFI).
The ionosphere reflects and refracts radio waves at low frequencies.
It inhibits transmissions from space below the ionospheric cut-off frequency. 
The cut-off frequency varies depending on the time of day and the Sun’s activity, although it can go down to 10\,MHz \citep{Toledo-Redondo2012}.
Also, it is difficult to find or create a radio-quiet zone for astronomical observations.
This is due to the presence of RFIs caused by the intercontinental communication signals which are broadcast by using the reflective feature of the ionosphere, which in turn makes observations at these radio frequencies complex \citep{Bentum2016ieee}. 

A space or moon based radio telescope can tackle the aforementioned challenges of such low frequency radio observations. 
There have been a few space missions dedicated to observations at these radio frequencies.
The first space mission at these radio frequencies was Radio Astronomy Explorer (RAE-1), made observations of the Galaxy's spectrum from 0.4 to 6.5\,MHz \citep{Alexander1969}.
A successor, RAE-2, was launched to the lunar orbit for measurements in the frequency range of 0.025 to 13MHz \citep{Alexander1975}.
Later, the Interplanetary Monitoring Platform\,(IMP-6) reported observations of galactic spectra at 22 frequencies, from 0.13 to 2.6\,MHz \citep{Brown1973}. 
Then the Netherlands Chinese Low-Frequency Explorer (NCLE), was launched in 2018, as part of China's Chang'E 4 Lunar mission. 
This is the most recent experiment for long-wavelength observations.
In addition, missions like Cassini-RPWS\,(Radio and Plasma Wave Science) and STEREO\,(Solar TErrestrial RElations Observatory) WAVES were designed, to perform in-situ low frequency observations of Saturn's magnetosphere and Coronal mass ejections from the Sun respectively \citep{Cecconi2007}. 

Another major development in the past decade is the implementation of the space-based very long baseline interferomenters VSOP/HALCA by Japan and RadioAstron by Russia \citep{Gurvits2020}.
Furthermore, there were numerous proposed concepts for space-based observations and instrumentation such as Farside Array for Radio Science Investigations of the Dark ages and Exoplanets\,(FARSIDE) \citep{Burns2019}, Orbiting Low Frequency Antennas for Radio Astronomy\,(OLFAR) \citep{Bentum2020, Bentum2011}, Distributed Aperture Array for Radio Astronomy In Space\,(DARIS) \citep{Bentum2011}, Space-based Ultra-long wavelength Radio Observatory\,(SURO) \citep{Baan2013}, and Formation-flying sub-Ionospheric Radio astronomy Science and Technology\,(FIRST) Explorer \citep{Bergman2009, Bentum2011}.
Also, there are a few projects under development such as America-led Dark Ages Polarimeter Pathfinder\,(DAPPER) for the frequency range 17 to 38\,MHz \citep{Burns2019AAS} and India-led Space Electric and Magnetic Sensor\,(SEAMS) for the frequency range 0.3 to 16\,MHz \citep{Borade2021, Borade2018}.

RFI suppression, array pattern\,(non-redundant baselines), high time resolution, time synchronisation, array element localization, antenna design, data handling, and space qualified instrumentation are some of the technological hurdles, in deploying the space based array \citep{Weiler2000}. 
This has led to the development of single satellite missions. 
Single satellite missions are the most often funded space-based missions due to their low technical complexity as well as budget constraints \citep{Lazio2020, Weiler2000, Shkolnik2018}. 
For very low frequency astronomical observation, it is necessary to localise and characterise\,(in terms of its polarisation property) the emissions from various sources to understand the emission mechanism of the source \citep{Lecacheux1978}.
This can be performed by a space based radio telescope array by use of the triangulation method \citep{Tools} however, a single radio telescope will require co-located antenna configurations and gonio-polarimetric methods\,(DoA methods).
The advances in antenna design and DoA methods/algorithms may overcome technological and economical obstacles. 
The widely known space projects that employ gonio-polarimetric methods for source localization are Cassini-RPWS, STEREO/WAVES, and NCLE \citep{Cecconi2007, Cecconi2005, Chen2010}. 
DoA estimation, is a technique for calculating the direction of an electromagnetic wave that is strongly reliant on the sensors'\,(antennas) orientation \citep{Rucker1997}.

The antenna array methods are mostly used in the development of methods for DoA estimation \citep{Nehorai1994}. 
Multi-Signal Classification\,(MUSIC), Estimation of Signal
Parameter via Rotational Invariance Technique\,(ESPRIT), Root-MUSIC and Modified-ESPRIT are among the widely used antenna array method based DoA estimators \citep{Waweru2014, Roy1986, Schmidt1986}.
These approaches are based on the eigen mode decomposition method wherein, a co-variance matrix is generated utilizing spatial smoothing to attain a full rank case. 
As a result, the DoA is well resolved while the technique is computationally complex \citep{Yilmazer2006}.
Thus, these techniques requires high computational resources which is difficult in case of in-situ computation.
So, techniques with low computational complexity with adequate resolution is essential.
Pseudo-vector based DoA, Analytical inversion method, and MPM based DoA are a few methods with low computational complexity.
The Pseudo-vector based DoA method estimates the EM wave direction by calculating a pseudo-vector using the spectral density tensor. 
This method is developed for tri-axial linear antenna configuration, wherein the antennas are arranged orthogonal to each other \citep{Carozzi2000}.
The analytical inversion method is developed for the STEREO/WAVES and Cassini-RPWS space missions \citep{Cecconi2005}.
This method is analytically derived by utilizing the correlation between the co-located antennas \citep{Cecconi2005} and has low computational complexity.
The analytical inversion method is developed for a specific antenna orientation as well as for in-situ observations \citep{Cecconi2007}.
The Matrix Pencil Method based DoA\,(MPM DoA) is another method for estimating the DoA which directly operates on the obtained spectrum without estimating the co-variance matrix depending on array configuration \citep{Yilmazer2006, Sarkar1995}.
This makes the MPM DoA method computationally light \citep{Yilmazer2006}.
The MPM DoA technique, which needs a triaxial antenna design, finds the DoA of multiple waves falling on the antenna arrangement using a unitary transformation approach and the LSM \citep{Daldorff2009, Chen2010}.

In this paper we propose a Snapshot averaged MPM\,(SAM) DoA algorithm, an improved version of the MPM DoA algorithm towards SEAMS mission\,(Section \ref{SEAMS}) along with the preliminary results of a proof of concept experiment performed for the DoA estimation.
The SAM DoA algorithm introduces an averaging and polarisation detection method based on the boundary conditions and orientation of the antenna structure in space.
This results in an increase in incoherent wave detection simultaneously increasing the accuracy and capability to differentiate between different types of polarisation.
This modification provides a way to study the polarisation characteristics of the emissions mechanisms at these radio frequencies bands.
To test the practicality of the results from the DoA algorithm, a laboratory experiment at the resonant frequency of the antenna\,($\sim72$\,MHz) has been performed.
In addition, this polarisation detection capability can aid in several different remote sensing applications like synthetic aperture radar\,(SAR) and Vegetation monitoring \citep{egido2012_RS,dvorsky2020_IEEE}. 

The current manuscript is organized in such a way that the Section \ref{SEAMS}, provides an overview of the SEAMS mission.
Section \ref{theory}, briefs about the principle of the EM wave direction and polarization detection. 
Section \ref{algo} and \ref{sim_set}, present the DoA algorithm description along with the proposed modification and the simulation setup, respectively. 
Section \ref{Results} describes the analysis and results of the Snapshot averaged Matrix Pencil Method\,(SAM) DoA algorithm and section \ref{scaled_exp} demonstrates it at the resonant frequency of the antenna. Section \ref{Conclusion} gives the Conclusion.

\section{Space Electric and Magnetic Sensor(SEAMS)}\label{SEAMS}

SEAMS is a Radio telescope which is currently being designed to
operate from 300\,kHz to 16\,MHz. 
The telescope will have three orthogonal electric and magnetic field sensors on-board. 
The first phase of the project is under development in SP Pune Uiversity and Phase II will follow.

In the first phase, only electric field vector sensors will be used for the measurement of RFI in low earth orbit. 
The system mainly consists of two orthogonal monopole antennas (electric field vector sensor, EFVS)  as a proof of concept and as a precursor to Phase II, RF front end with matching network, filters and gain stages for both arms and a two channel data acquisition and analysis system with Telemetry-Telecommand interface.
The first phase will be deployed in the Low Earth Orbit (LEO) on the 4th stage of the ISRO-PSLV rocket with the objectives of analysing the acquired RFI from the Earth, detect Auroral Kilometric Radiation (AKR), lightning in the atmosphere, strong solar bursts, etc.
This phase will also provide insight about the feasibility of using Commercial Off-The-Shelf\,(COTS) components to design payloads in LEO and to reduce production cost and up-gradation required for SEAMS phase-2.

The SEAMS Phase-2, will have three orthogonal electric and magnetic sensors and the payload will be placed on the far side of the Moon or in the Moon-Earth L2 to avoid the RFI from Earth.
The details of the science goals for Phase II are evolving but are described in the upcoming SEAMS- Phase\,I article (Kulakrni, A., etal 2022, under preparation).

\section{DoA Estimation Method}\label{theory}
Analysis of the spectral density tensor\,($S = \Vec{E}\Vec{E}^\dagger$) can provide the direction and polarisation of the EM wave based on the field vector in Eq (\ref{eqn:eqn1}).
Based on the Gell-Mann \textit{SU(3)} matrix, the anti-symmetric part of $S$ can be converted into a dual pseudo-vector containing information about the EM wave's direction and polarization\citep{Carozzi2000}.

\begin{equation}
    \label{eqn:eqn1}
    \Vec{E} = e_x \hat{a_x} + e_y \hat{a_y} + e_z \hat{a_z}
\end{equation}

The pseudo-vector\,($\Vec{V}$) associated with the spectral density tensor\,(Eq. (\ref{eqn:eqn2})) is a parallel vector to the wave vector $k$ and represents a three dimensional analogy of Stokes parameter $V$ (of the $I, Q, U$, and $V$)  \citep{Carozzi2000, Chen2010, Harsha2021ieee}. 
\begin{equation}
    \label{eqn:eqn2}
    \Vec{V} = -2Im[e_y e_z^* \hat{a_x} - e_x e_z^* \hat{a_y} + e_x e_y^* \hat{a_z}]
\end{equation}

Thus, the DoA of the EM wave\,(i.e, azimuth and elevation angle) is determined in the spherical coordinate system\,(See Eq. (\ref{eqn:eqn3})). 
The polarization of the EM wave can be estimated by normalizing the magnitude of $\Vec{V}$ with respect to the three dimensional analogy of the Stokes $I$ parameter\,(i.e, $polarization = \lvert\Vec{V}\rvert/I$ where, $I = e_xe_x^* + e_ye_y^* + e_ze_z^*$) \citep{Daldorff2009}.  
This results in a value between 0 to 1 where, 0 signifies completely linear polarization\,(LP) and 1 signifies completely circular polarization\,(CP). 
\begin{equation}
    \label{eqn:eqn3}
    \Vec{V} = v\cdot(sin\theta cos\phi\ \hat{a_x} + sin\theta sin\phi\ \hat{a_y} + cos\theta\ \hat{a_z})
\end{equation}
Here, $\theta$ and $\phi$ represent the Elevation and Azimuth angles and can be rewritten as \citep{Chen2010}
\begin{equation}
    \label{eqn:eqn4}
    \begin{aligned}
        \theta &= arccos({V_z / \lvert\Vec{V}\rvert})\\
        \phi &= 
        \begin{cases}
            \arctan(V_y / V_x) + \pi/2,     & \text{if } V_x< 0\\
            \arctan(V_y / V_x),             & \text{if } V_x> 0
        \end{cases}
    \end{aligned}
\end{equation}

\subsection{Algorithm description}\label{algo}

\begin{figure}[h!]
    \centering
    \includegraphics[width=0.6\linewidth]{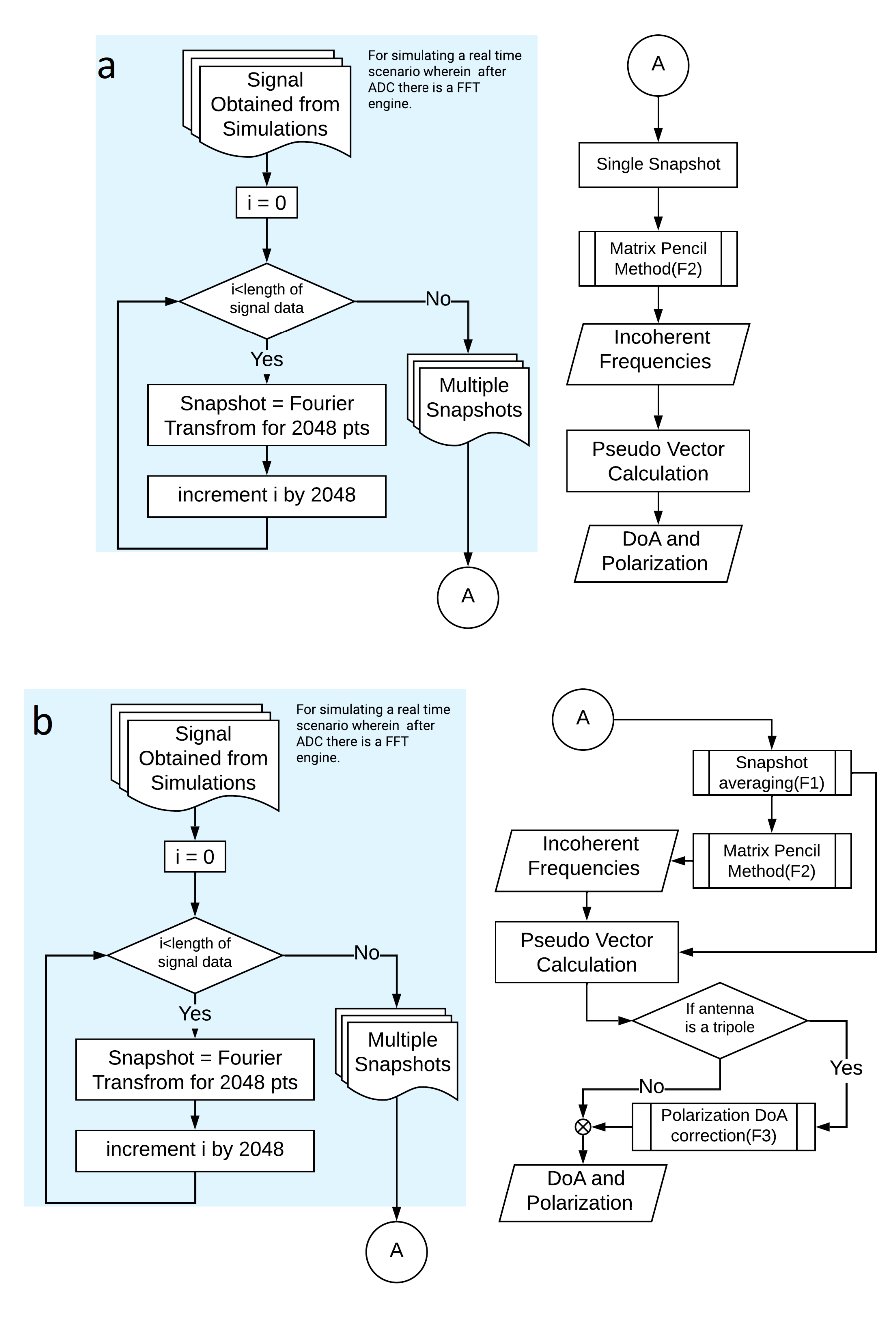}
    \caption{Estimation Algorithm Flowchart - (a) MPM DoA algorithm, (b) Modified-MPM DoA algorithm. Both the algorithm use Matrix pencil method\,(F2) \citep{Sarkar1995} to estimate the number of incoherent waves and angular frequency \citep{Daldorff2009, Chen2010}. In the modified algorithm\,(shown in (b)) two modifications are introduced (1) Snapshot averaging algorithm\,(F1) and (2) Polarization, DoA correction\,(F3).}
    \label{fig:Fig1}
\end{figure}

A single element of a tri-dipole or a tripole receiving $N$ incoherent EM waves is expressed as
\begin{equation}
    \begin{aligned}
        s_{x'}(t)=\sum_{n=1}^{N} C_{x'}^n e^{i\theta_{x'}^n} e^{i\omega^n t} + n(t)
    \end{aligned}
\end{equation}
where, $C_{x'}$, $\theta_{x'}$, and $\omega$ are the amplitude, phase and frequency of the incident incoherent EM wave on the antenna along the $x'$ axis which is aligned at an angle to the reference axis [Sec. \ref{sim_set}]. Then the sampled signal will be represented as
\begin{equation}
    \begin{aligned}
    \label{eqn:eqn6}
         S_{x'}[k] = \sum_{n=1}^{N} (C_{x'}^n e^{i\theta_{x'}^n} e^{i\omega^n t_0}) e^{i\omega^n k\delta} + n(t_0 + k\delta)
    \end{aligned}
\end{equation}
where, $S_{x'}[k]$ is a discretized signal containing $N$ incoherent EM waves, $(t_0 + k\delta)$ represents the sampling time, $n(t_0 + k\delta)$ is noise, $\delta$ is the sampling period, and $\omega^{n}$ represents the angular frequency of the $n^{th}$ incoherent wave. 
To compute the complex amplitude of the received EM wave and obtain the DoA using Eq. (\ref{eqn:eqn2}), the best estimations of $N$ and $\omega^n$ must be found.
Therefore, the developed algorithm estimates the frequency of emission of the incident incoherent wave.
A flow chart of SAM DoA algorithm and its predecessor is shown in figure \ref{fig:Fig1}.
The following is the description of the developed algorithm:
\begin{itemize}
    \item The acquired signal from each antenna is averaged for n snapshots by calculating the average value of the phase difference between the antennas.  
    \item Thereafter, to reduce the computational cost a beam forming addition \citep{Chen2010} of snapshot averaged signal is implemented.
    \item In order to estimate the $N$ incoherent frequencies MPM \citep{Sarkar1995,Yilmazer2006, Daldorff2009, Chen2010} is used and is summarised in Appendix  \ref{secA2}.
    \item The N incoherent frequencies $\omega^n$ obtained from the MPM are then used to find the complex amplitudes for each axis using a constrained LSM. Wherein, the prior phase information from the averaging method is provided to the constrained LSM. The least square method calculates the $min(\lvert KF_i - S_i\rvert^2)$ such that $Qp \leq F_i$ where, $K$ is matrix of dimension $M\times M$ as the signal consists of $M$ samples\,($M$ = $0,1,2,...,M-1$), $F_i$ is the complex amplitude matrix, $S_i$ is the spectral density matrix, $p$ is the prior phase information from averaging method, and $Q$ is the amplitude matrix.
    \item Then the DoA is calculated using the pseudovector formulation $\Vec{V} = -2\{A_yA_zsin(\delta_y-\delta_z)\hat{a_x}-A_xA_zsin(\delta_x-\delta_z)\hat{a_y}+A_yA_zsin(\delta_x-\delta_y)\hat{a_z}\}$ and by considering equations (\ref{eqn:eqn2}) and (\ref{eqn:eqn4}), as well as the detection of polarization ($P = \lvert\Vec{V}\rvert/I$) and the plane wave direction constraints in our simulation setup (i.e., azimuth $[0^\circ,360^\circ)$ and elevation $[0^\circ,90^\circ]$). There is only one case where $V_z$ is negative: when the incident wave has a LHCP. 
    Since polarisation is defined as $P = \lvert\Vec{V}\rvert/I$, the $P$ range is limited to $[0, 1]$.
    As a result, the method will be limited to distinguishing between LP and CP.
    The polarisation finding equation is updated to Eq. (\ref{eqn:eqn12}) to increase the algorithm's polarisation detecting capacity.
    \begin{equation}
        \label{eqn:eqn12}
        \begin{aligned}
            P &= 
            \begin{cases}
                \lvert\Vec{V}\rvert/I,             & \text{if } V_z> 0\\
                -\lvert\Vec{V}\rvert/I,     & \text{if } V_z< 0
            \end{cases}
        \end{aligned}    
    \end{equation}
\end{itemize}

\section{Antenna and Simulation Setup}\label{sim_set}
 
\begin{figure}[h]  
    \centering
    \includegraphics[width=1\linewidth]{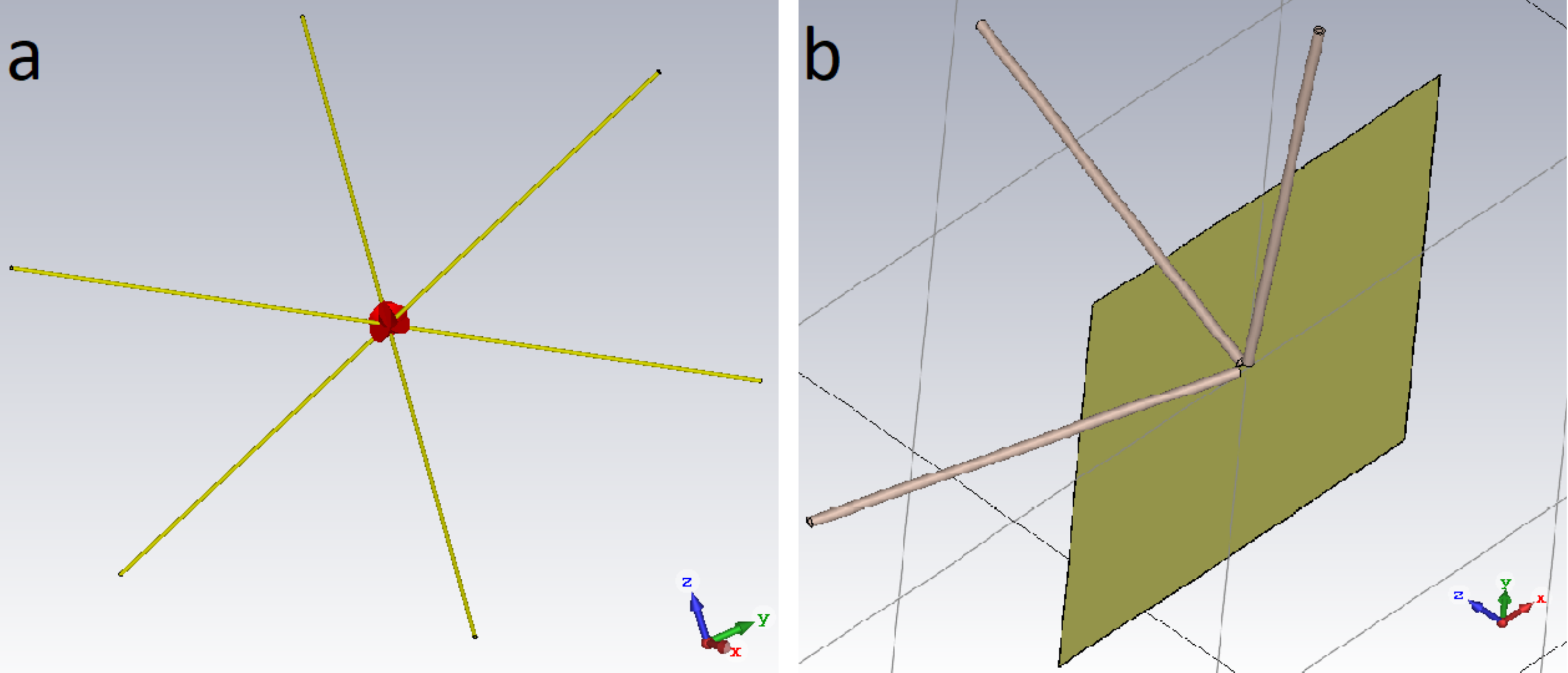}
    \caption{Antenna configurations - (a) Tri-Dipole - 3 orthogonal co-located dipoles and (b) Tripole - 3 orthogonal co-located monopoles}
    \label{fig:Fig2}
\end{figure}

Two types of antenna configurations were simulated for the testing of the DOA algorithm in relation to the SEAMS mission: (1)\,triaxial dipoles\,(Tri-dipole) and (2)\,triaxial monopoles\,(Tripole). 
In the Tri-dipole and Tripole configurations, the dipoles and monopoles are $\sim$2\,m and $\sim$1\,m long, respectively.
The length of the antenna should be adjusted according to the active matching network's sensitivity or operating frequency \citep{Nordholt1980}.
As illustrated in Fig. \ref{fig:Fig2}, the triaxial dipole and monopole are made up of three mutually orthogonal co-located dipoles and monopoles. 
The dipoles or monopoles are orientated along the $x'$, $y'$, and $z'$ axes, with each subtending an angle of $35.3^\circ$ \footnote{$35.3^\circ$ is the angle between the each monopole and the ground plane. This is derived by equating all the direction cosines of the monopoles}.
Equation (\ref{eqn:eqn0}) describes the connection between the unit vectors associated with the antenna axes ($x'$, $y'$, and $z'$, from the antenna frame) and the frame axes or global axes (from the reference frame) \footnote{Here the coordinate setting is such that each orthogonal monopoles subtends an angle of $35.3^\circ$ with the ground plane\,(x-y plane) and one can consider the new locations of the monopoles as the new local coordinates; then the relationship between the local and the reference coordinate is given by equation \ref{eqn:eqn0}.}.

\begin{equation}
    \label{eqn:eqn0}
    \begin{aligned}
    \hat{a_x} &= \frac{\hat{a_{x'}}}{\sqrt{3}} - \frac{\hat{(3+\sqrt{3})a_{y'}}}{6} + \frac{\hat{(3-\sqrt{3})a_{y'}}}{6} \\
    \hat{a_y} &= \frac{\hat{a_{x'}}}{\sqrt{3}} + \frac{\hat{(3-\sqrt{3})a_{y'}}}{6} - \frac{\hat{(3+\sqrt{3})a_{y'}}}{6} \\
    \hat{a_z} &= \frac{\hat{a_{x'}}}{\sqrt{3}} + \frac{\hat{a_{x'}}}{\sqrt{3}} + \frac{\hat{a_{x'}}}{\sqrt{3}}
    \end{aligned}
\end{equation}

The field vector $\Vec{E}$\,(Eq. (\ref{eqn:eqn1})) is characterised in the reference frame axes according to the antenna orientation, and the field vector components in the reference frame are recast in terms of the field vector components in the antenna frame using Eq. (\ref{eqn:eqn0}). 

\begin{equation}
     \label{eqn:eqn5}
     \begin{aligned}
         e_x &= \frac{E_{x'}}{\sqrt{3}}-\frac{E_{y'}(3+\sqrt{3})}{6}+\frac{E_{z'}(3-\sqrt{3})}{6} \\ 
         e_y &= \frac{E_{x'}}{\sqrt{3}}+\frac{E_{y'}(3-\sqrt{3})}{6}-\frac{E_{z'}(3+\sqrt{3})}{6} \\ 
         e_z &= \frac{E_{x'}}{\sqrt{3}}+\frac{E_{y'}}{\sqrt{3}}+\frac{E_{z'}}{\sqrt{3}}
     \end{aligned}
\end{equation}

where, $e_x$, $e_y$ and $e_z$ is the field vector components in reference frame, and $E_{x'}$, $E_{y'}$, and $E_{z'}$ are the field vector components in the antenna frame \citep{Harsha2021ieee}. 

In addition, the antenna configuration is designed and simulated using the CST antenna design software with a plane wave as an excitation source. 
This is analogous to a free space environment with no RFIs or noise. 
The simulation was carried out for plane waves arriving from various directions at different frequencies in the sub 20 MHz band (300\,kHz - 16\,MHz) and with varying polarizations. 
As the SEAMS mission will be deployed on the far side of the moon, a side of the antenna will always face the moon. 
Despite the fact that the satellite will be on the far side of the moon, low-level RFIs and reflections from the moon's surface will cause noise in the system \citep{Bentum2020, Bentum2019}.
If the tripole antenna always faces the sky with its ground towards the moon, it can lessen the effect of the noise and RFI.
Thus, the plane wave direction in the simulation is constrained to the azimuth - $[0^\circ,360^\circ)$ and elevation - $[0^\circ,90^\circ]$ since the moon is on one side of the antenna arrangement with the other side facing space.

\section{Simulation Results and Discussion}\label{Results}

As discussed in the above section\,(\ref{sim_set}), the antenna configuration simulation is performed for 1000 trials with RHCP and LHCP. 
The data set generated by the antenna simulation is of unique frequency. 
To simulate multiple incoherent frequencies, noise of diffferent Signal to Noise Ratio\,(SNR) values\,(1 to 30\,dB) is added to the time domain data.
The performance of the DoA algorithm is evaluated using the following criteria: 
\begin{enumerate}
    \item Singular Value Ratio\,(SVR)\footnote{The SVR factor is the summation of the ratio between the consecutive values of the eigen values obtained from Singular value decomposition\,(SVD). 
    This is a good measure to test the performance of the algorithm because the order of eigen values in the SVD are arranged from the most prominent feature in the signal to least prominent feature. 
    If there are a few incoherent waves incident then the change observed in the consecutive eigen values will be abrupt or steep. If the ratio between the consecutive eigenvalues is high, it means that the signal can be detected easily.} response with change in SNR,
    \item SVR response with change in number of incoherent sources.
    \item Root Mean Square Error\,(RMSE\,=\,$\sqrt{N^{-1}\Sigma_{i=1}^N (x_{estimated}-x_{actual})^2}$) of azimuth and elevation angles with change in SNR,
    \item RMSE of azimuth and elevation angles with change in $N$, and
    \item A polarization detection table.
\end{enumerate}

\begin{figure}[!ht]
    \centering
    \includegraphics[width=1\textwidth]{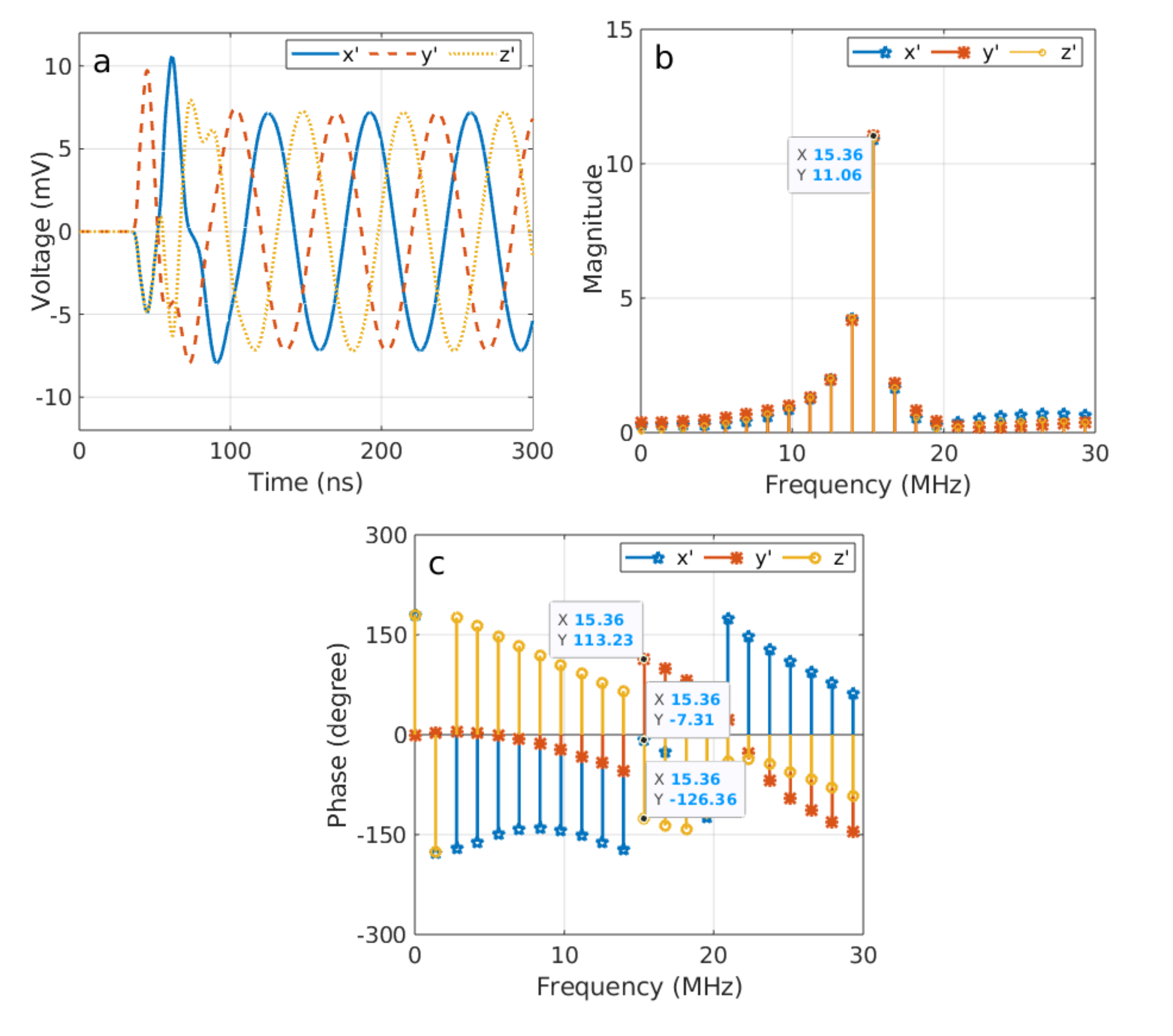}
    \caption{(a) Time domain Voltage readings, (b) Magnitude and (c) Phase of the Frequency spectrum of the CST simulation of antenna structure shown in Fig. \ref{fig:Fig2} for RHCP plane wave excitation of 15\,MHz.}
    \label{fig:Fig3}
\end{figure}

In order to evaluate the algorithm, first a CST simulation was carried out using a plane wave excitation and the output voltage from the simulations are recorded.
The simulation was performed at high time resolution such that the difference in the phases and amplitudes of the received signal is clearly visible when plotting time domain data as shown in Fig. \ref{fig:Fig3}(a); Fig. \ref{fig:Fig3}(b) and \ref{fig:Fig3}(c) show the Amplitude and phase of the signal in the frequency domain.
The simulation results shown in Fig. \ref{fig:Fig3} were obtained for RHCP plane wave with frequency 15\,MHz approaching from $+z$ direction.
The propagation vector of the plane wave will subtend an equal angle with all the antennas of the tripole and the tri-dipole given our antenna orientation\,(Section \ref{sim_set}).
Thus, the voltage induced on each antenna should be $120^\circ$ out of phase with each other, and the same is observed from the simulation as shown in Fig. \ref{fig:Fig3}(a). 
This phase difference is more distinct in the Fig. \ref{fig:Fig3}(c) - phase of the Fourier spectrum.
Furthermore, the signal received by the both the antenna configurations is observed to be the same.

The data obtained from the simulation are then contaminated with noise.
Thereafter the noisy signal is used in the DoA algorithm to find the direction and polarization of the incident wave. 
As discussed in the section\,(\ref{algo}), $N$ and $\omega^n$ should be estimated first using the MPM. 
To evaluate the efficiency of the algorithm SVR is determined, which is the ratio of the eigen values obtained by decomposing $\Gamma_R$\,($\sigma_{N}/\sigma_{N+1}$) \citep{Chen2010, Kanjilal1995}.
This singular value obtained as a result of decomposition of the $\Gamma_R$ will gradually decrease depending on the number of incident waves.
Thus, as per the definition of SVR, the higher the value of the SVR, the better is the estimation of $N$ and $\omega^n$ \citep{Kanjilal1995, Chen2010}.

\begin{figure}[!ht]
    \centering
    \includegraphics[width=1\textwidth]{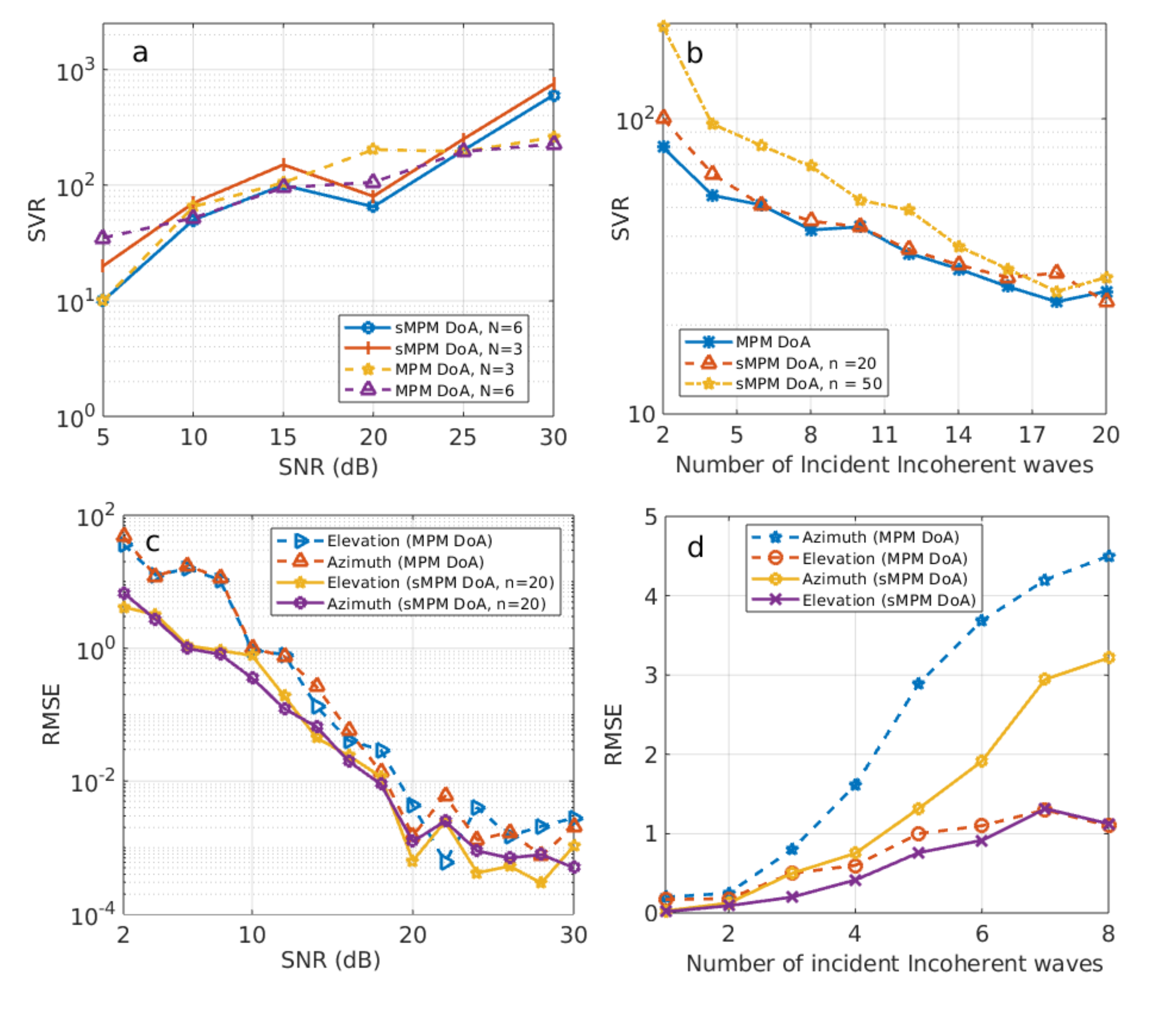}
    \caption{Comparing the performance of the MPM DoA and the proposed SAM DoA estimation. (a)\,Effect of variation of the SNR on the SVR, a measure of efficiency of the DoA algorithm when estimating $N$ and $\omega^n$ if the number of incoherent wave incident on the antenna configuration is $N = 3$ and $N = 6$; (b)\,Effect of the change in the incident incoherent wave on the SVR at SNR\,=\,15dB and $n = 20$; (c)\,Effect of the SNR on the detection of the direction of the Radio wave by estimating RMSE and the number of snapshots averaged in the SAM DoA estimation is $n=20$; (d)\,Effect of the change in incident incoherent wave on the detection of direction of the Radio wave with $n=20$ in the SAM DoA estimation and SNR\,=\,15\,dB.}
    \label{fig:Fig4}
\end{figure}

Figure \ref{fig:Fig4}(a) and \ref{fig:Fig4}(b) shows the effect of the change in SNR and $N$ on SVR response of the SAM DoA algorithm and MPM DoA algorithm. 
It is observed that the SVR value increases with the SNR value which is as expected.
It is also noticed from Fig. \ref{fig:Fig4}(a) and \ref{fig:Fig4}(b) that the increase in the incoherent waves or frequency reduces the SVR value. 
This indicates that the detection of an incident incoherent wave would be difficult for multiple incident waves. 
Furthermore, it is observed that the SAM DoA algorithm improves the SVR response with respect to change in the SNR along with an increase in the number of incoherent waves.
From Fig. \ref{fig:Fig4}(b), it is observed that increasing the number of snapshot averages improves the SVR value. 
Thus, the SAM DoA method provides flexibility to increase the SVR value by averaging a large number of snapshots. 

\begin{figure}[!ht]
    \centering
    \includegraphics[width=1\textwidth]{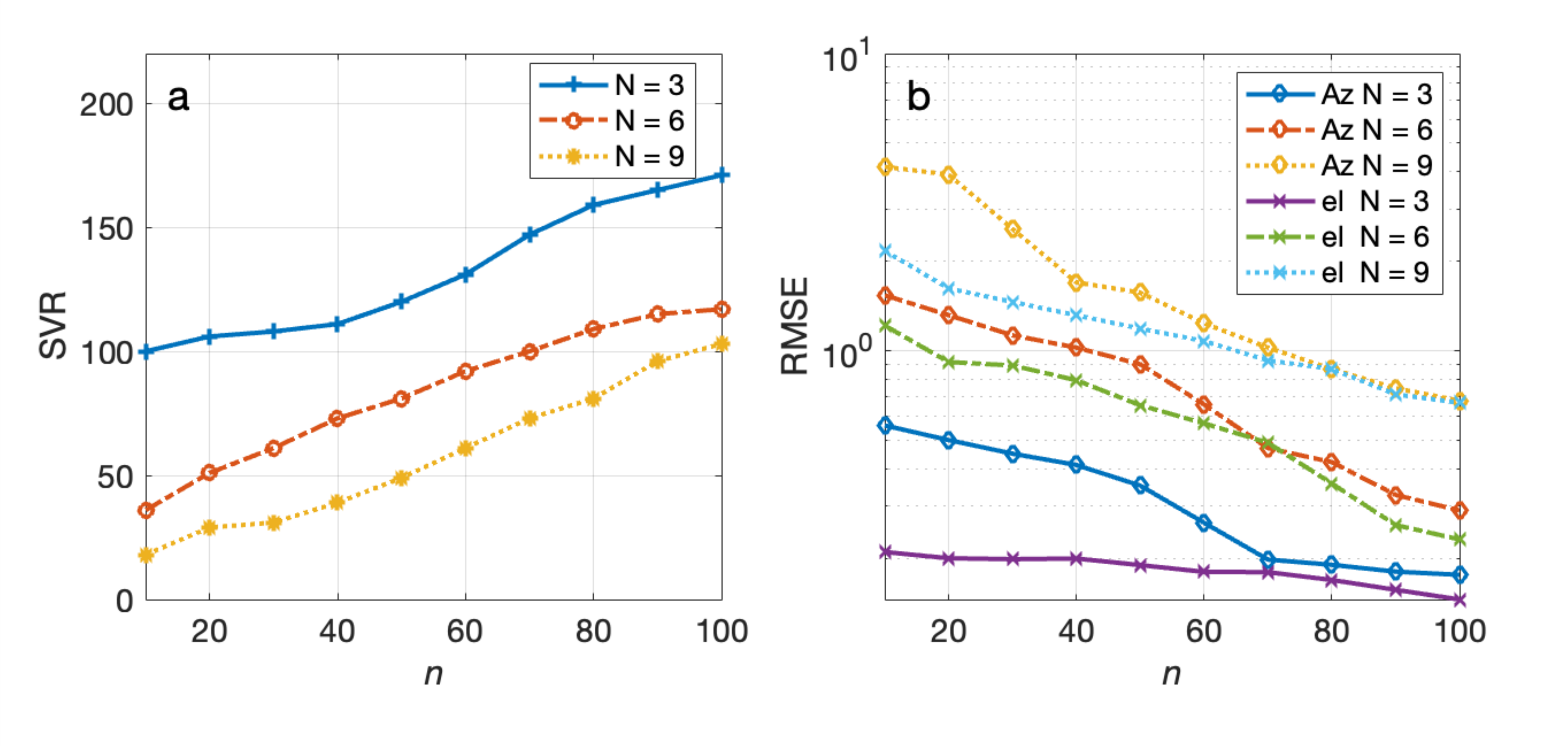}
    \caption{Response of SAM DoA estimation with change in number of incoherent wave\,($N$) and number of averaged snapshot\,($n$). (a)\,SVR response with change in the $n$; (b)\,RMSE response with respect to $n$.}
    \label{fig:Fig5}
\end{figure}

\begin{figure}[!ht]
    \centering
    \includegraphics[width=1\linewidth]{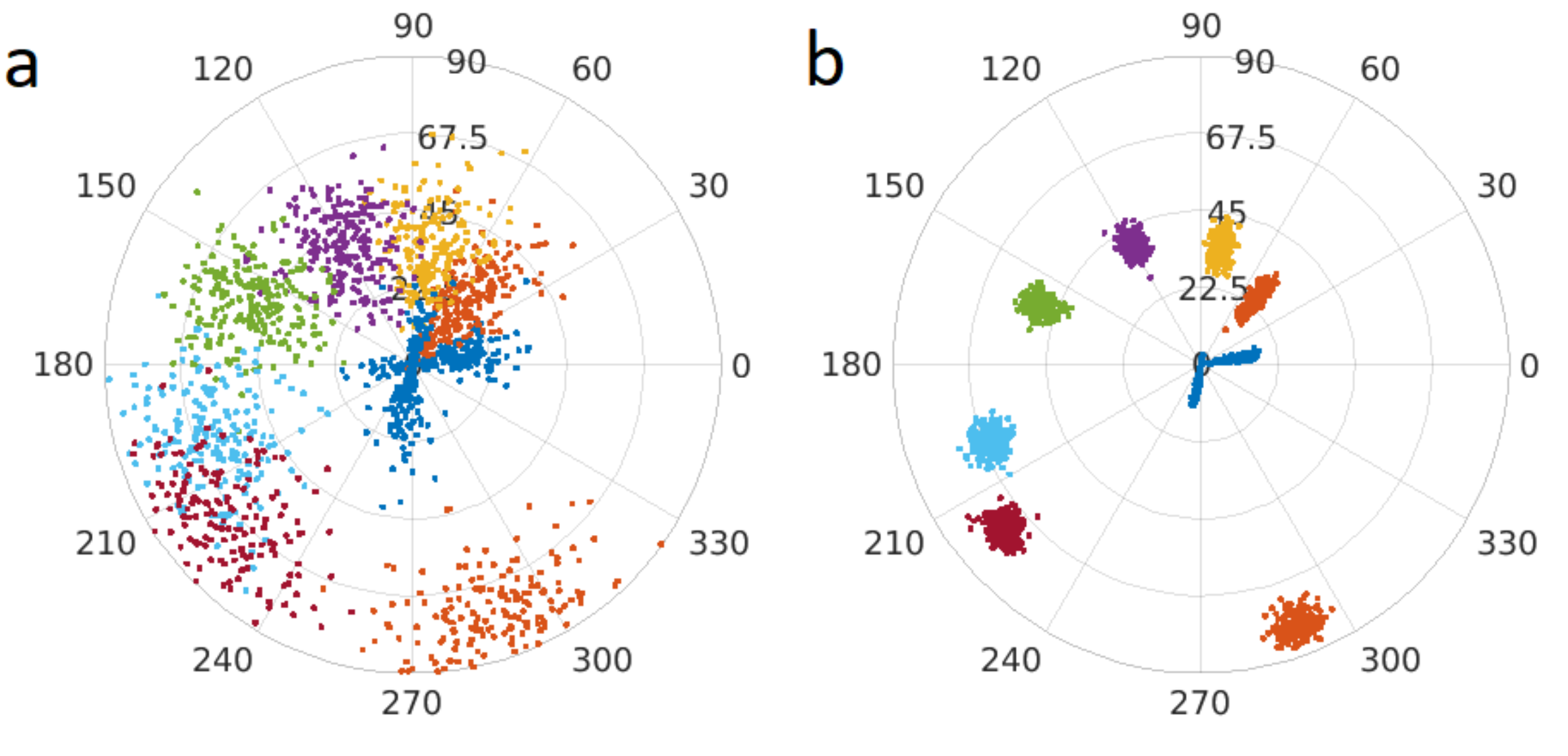}
    \caption{Polar representation Azimuth and elevation angles of the estimated DoA for 9 incoherent frequencies incident on the antenna simultaneously from different directions of the radio sources at SNR of 15\,dB. (a) shows the DoA plot for MPM DoA method and (b) shows the plot for SAM DoA method for $n=50$. }
    \label{fig:Fig6}
\end{figure}

The accuracy of detecting the direction and the polarization is shown in the Fig. \ref{fig:Fig4}(c), \ref{fig:Fig4}(d), and Table \ref{tab:pol_dec}.
As discussed in the above section\,(\ref{sim_set}), the simulation is only performed for the plane wave direction in the range of azimuth $0^\circ$ to $360^\circ$ and elevation $0^\circ$ to $90^\circ$. 
In this case it observed that the RMSE decreases when SNR increases, as can be expected.
It is also observed that the error in the direction estimation increases with an increase in the incoherent frequencies in the signal. 
However, it is to be noted that the RMSE of the SAM DoA algorithm has a significant improvement for a large number of averaged snapshots. 
Hence, if the number of snapshots to be averaged is increased then the detection of the direction would improve.
This improvement in detection can also be observed by Fig. \ref{fig:Fig5}, representing the response of SVR and RMSE with respect to increase in the number of averaged snapshots.
In the figure the single RMSE value or the effective RMSE$_{eff}$ for multiple incident incoherent wave is calculated by RMSE$_{eff}=\sqrt{\sum_{i=1}^N RMSE_i}$ where, RMSE$_i$ is error due to individual waves.
Also the improvements mention can also be observed in Fig. \ref{fig:Fig6} wherein, polar plot images for 9 incoherent sources having azimuth along the perimeter and elevation along the concentric circles.
Fig. \ref{fig:Fig6}(b) shows the polar plot of the estimated DoA for SAM DoA are shown for SNR\,=\,15\,dB for 9 incoherent frequencies from different directions with number of snapshot averaged is 50.
By comparing the images in Fig. \ref{fig:Fig6}(a) and \ref{fig:Fig6}(b) it is noticed that the area of the estimated DoA is reduced.

\begin{table}[!ht]
    \centering
    
    \begin{tabular}{c|c|c|c|c}
    \hline
        $N$ & SNR & Actual Polarization & Detected Polarization & Detected Polarization\\
        & (in dB) & of the incident waves & (MPM DoA) & (SAM DoA)\\
        \hline
          &  & 1\,RHCP, & 2\,EP, and & 1\,RHEP,\\
         3 & 10 & 1\,LHCP, and & 1\,LP & 1\,LHEP, and\\
          &  &  1\,LP & &1\,LP\\
         \hline
          &  & 1\,RHCP, & 2\,CP, and & 1\,RHCP,\\
         3 & 30 & 1\,LHCP, and & 1\,LP & 1\,LHCP, and\\
          &  &  1\,LP & &1\,LP\\
         \hline
          &  & 2\,RHCP, & 8\,EP, and & 4\,RHEP\\
          &  & 2\,LHCP, and & 2\,LP & 4\,LHEP, and\\
         10 & 10 & 2\,RHEP, and & & 2\,LP\\
          &  & 2\,LHEP, and & & \\
          &  &  2\,LP & &\\
         \hline
           &  & 2\,RHCP, & 3\,CP, & 2\,RHCP,\\
          &  & 2\,LHCP, and & 5\,EP, and & 1\,LHCP,\\
         10 & 30 & 2\,RHEP, and & 2\,LP & 2\,RHEP,\\
          &  & 2\,LHEP, and & & 3\,LHEP, and\\
          &  &  2\,LP & &2\,LP\\
         \hline
    \end{tabular}
    \caption{Polarization detection table describing the classification of the detected polarization by the MPM DoA algorithm and SAM DoA algorithm. Here $N$ is the number of incident incoherent waves.}
    \label{tab:pol_dec}
\end{table}
Table \ref{tab:pol_dec} illustrates the improvement in the detection of the polarization due to the modified algorithm introduced by equation (\ref{eqn:eqn12}). 
From the table it is clearly observed that the SAM DoA algorithm is capable of differentiating different polarization.

\begin{table}[!ht]
    \centering
    \begin{tabular}{p{1.8cm}|p{1.5cm}|p{1.5cm}|p{1.5cm}|p{1.5cm}}
    \hline
        Features & Analytical Inversion Method & Pseudo vector estimation based DoA  & MPM DoA Method & SAM DoA Method \\ 
        \hline
        \footnotesize{Direction Estimation Error (SNR=$\sim15$\,dB)} & $<6^\circ$ & $>10^\circ$ & $<1^\circ$ & $<0.5^\circ$ \\
        \hline
        \footnotesize{Maximum Computational Complexity} & \footnotesize{O($5N^3log_2N+88N+20N^2$)} &  \footnotesize{O($3Nlog_2N+4N^3$)} & \footnotesize{O($3Nlog_2N+(N-L)(L+1)+(M-L)^2(2L+2)+(N-L)^4$)} & \footnotesize{O($3n^2Nlog_2N+(N-L)(L+1)+(M-L)^2(2L+2)+(N-L)^4+N(n^2+1)$)} \\
        \hline
        \footnotesize{Required Computational time } & 0.135s & 0.104s & 1.53s & 1.76s\\
        \hline
        \footnotesize{Polarization detection} & Yes  & Yes & Yes & Yes \\
        \hline
        \footnotesize{Differentiation between LH and RH polarization} &  No & No & No & Yes \\
        \hline
        \footnotesize{Space mission for which algorithm is Proposed} & \footnotesize{Cassini - RPWS and STEREO WAVES} & - & \footnotesize{NCLE} & \footnotesize{SEAMS}\\
        \hline
        \footnotesize{References} & \citep{Cecconi2005} & \citep{Carozzi2000} & \citep{Daldorff2009}, \citep{Chen2010}, \citep{Chen2018} & Current work \\
        \hline
    \end{tabular}
    \caption{Performance of SAM DoA with other proposed algorithms. Four algorithms are compared with the SAM DoA method with respect to its Estimation error, Computational complexity, Detection capability of polarization. The estimation error is reported for single source with SNR of 15\,dB. The approximate value of the Maximum computational complexity is calculated by referring to the algorithm description provided in the articles mentioned in the reference row. The required computational time is calculated for clock speed - 1 ns, matrix size N=300, pencil parameter L=100, and average points n = 20.}
    \label{tab:tab3}
\end{table}

Table \ref{tab:pol_dec} shows that the MPM DoA algorithm can only distinguish between Linear, Circular, and Elliptical Polarization, while the improved SAM DOA method can identify all forms of polarization.
Table \ref{tab:tab3} compares the computational complexity of various methods such as Analytical Inversion Method \citep{Cecconi2005}, Pseudo-vector based method \citep{Carozzi2000}, and MPM DoA method \citep{Daldorff2009, Chen2010}. 
The computational complexity is calculated by considering the involved processes like Fourier tranformations, correlations, SVD etc.
This is done by studying the formulation of these algorithms.
From table \ref{tab:tab3}, it is observed that the SAM DoA is computationally more complex and the accuracy of estimation is the highest for this method. 
In addition to this, the SAM DoA algorithm provides a flexible parameter $n$\,(i.e., average points), and increasing this parameter increases the accuracy of the estimation\,(Fig. \ref{fig:Fig4}) as well as the computational time.
This parameter is a loop parameter, and therefore it will not add to the algorithm's complexity but will affect the computing time.

\section{Setup and Result of a scaled experiment}\label{scaled_exp}

\begin{figure}[!ht]
    \centering
    \includegraphics[width=1\linewidth]{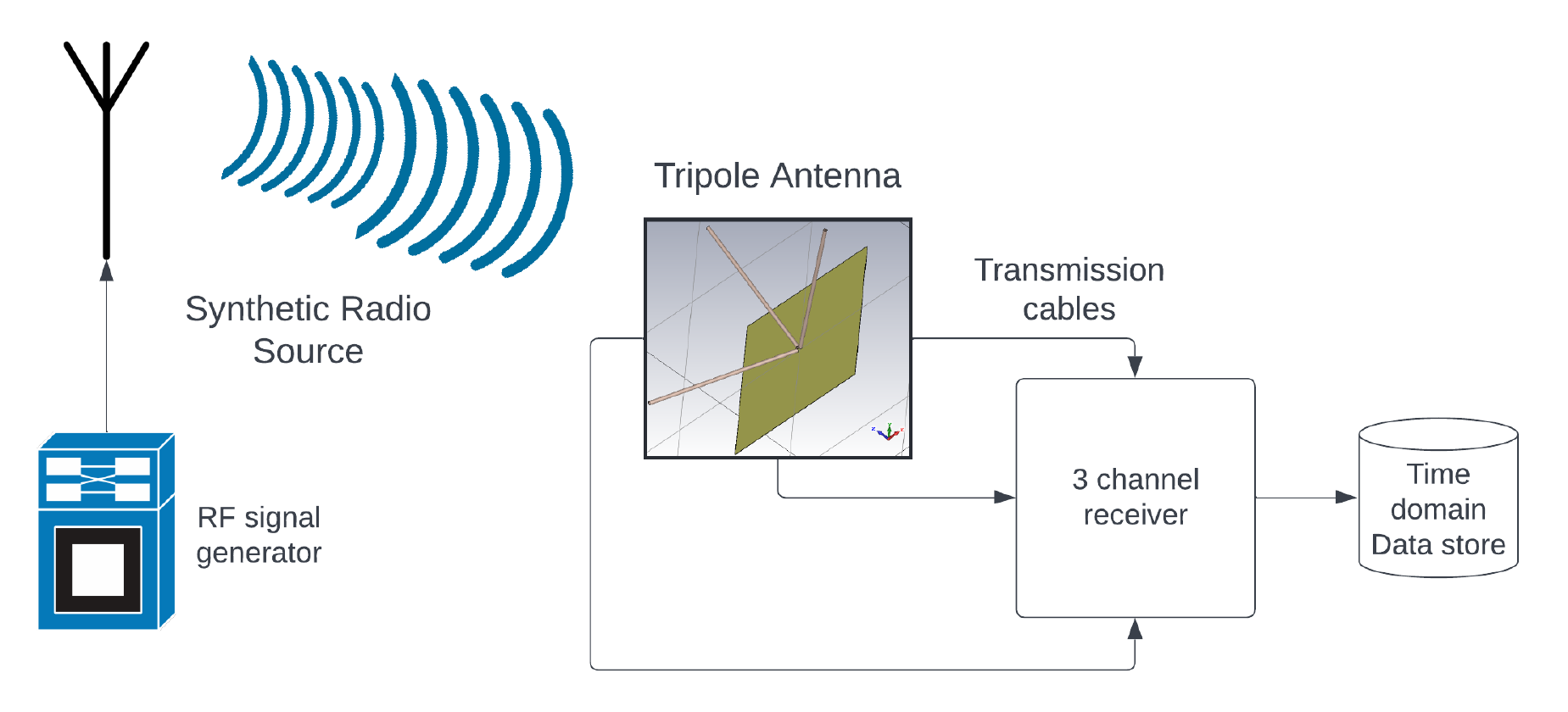}
    \caption{A Block diagram of the experimental setup. A high frequency oscilloscope\,(DSO - 100\,kHz to 2\,GHz) is used to record the time domain data in this setup.}
    \label{fig:Fig7}
\end{figure}

\begin{figure}[!ht]
    \centering
    \includegraphics[width=0.7\linewidth]{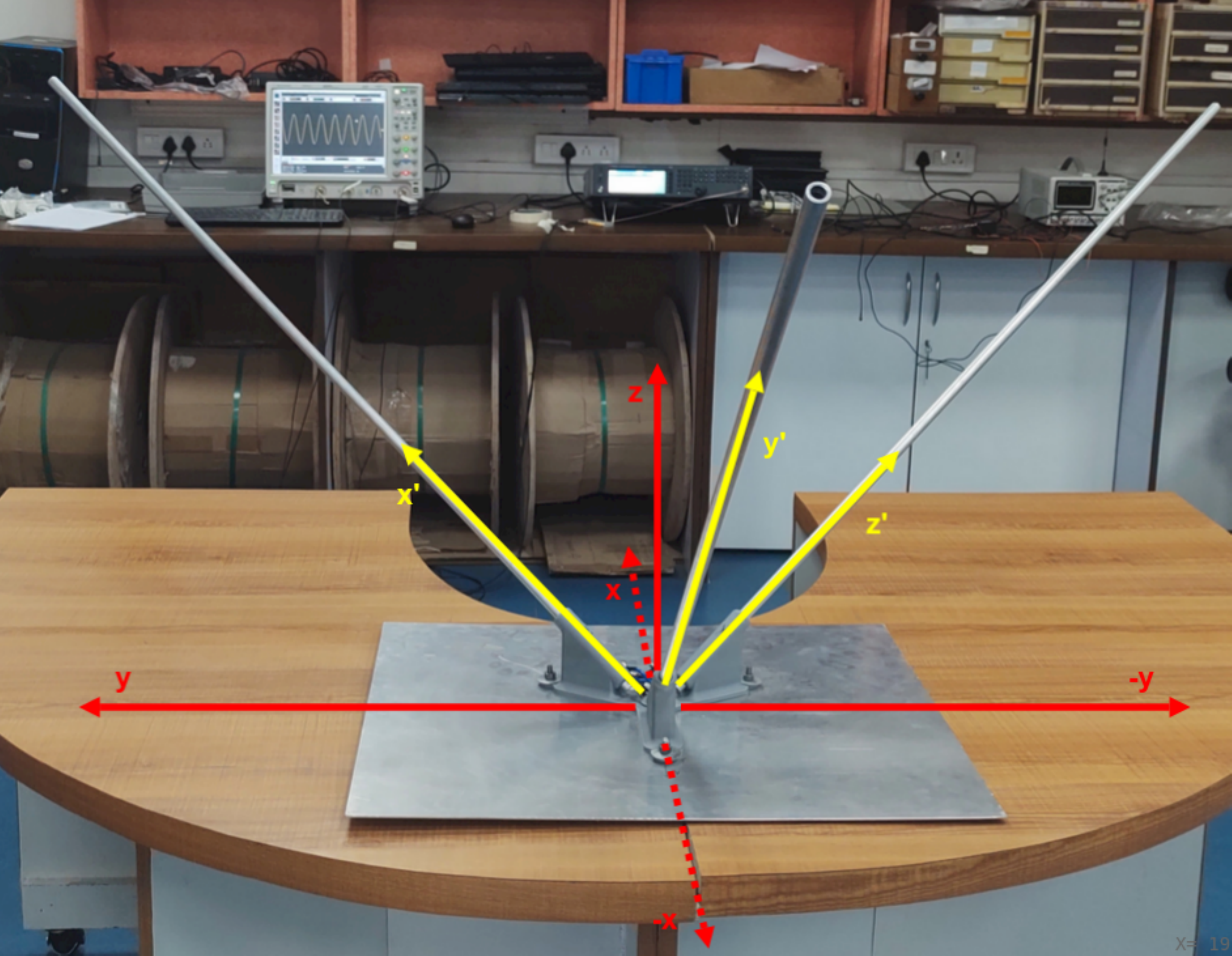}
    \caption{Fabricated tri-axial monopole resonant at $\sim$72\,MHz with orthogonal antenna axis $x',\ y',\ z'$ and reference plane axis $x,\ y\ ,z$. }
    \label{fig:Fig8}
\end{figure}

\begin{figure}[!ht]
    \centering
    \includegraphics[width=0.7\linewidth]{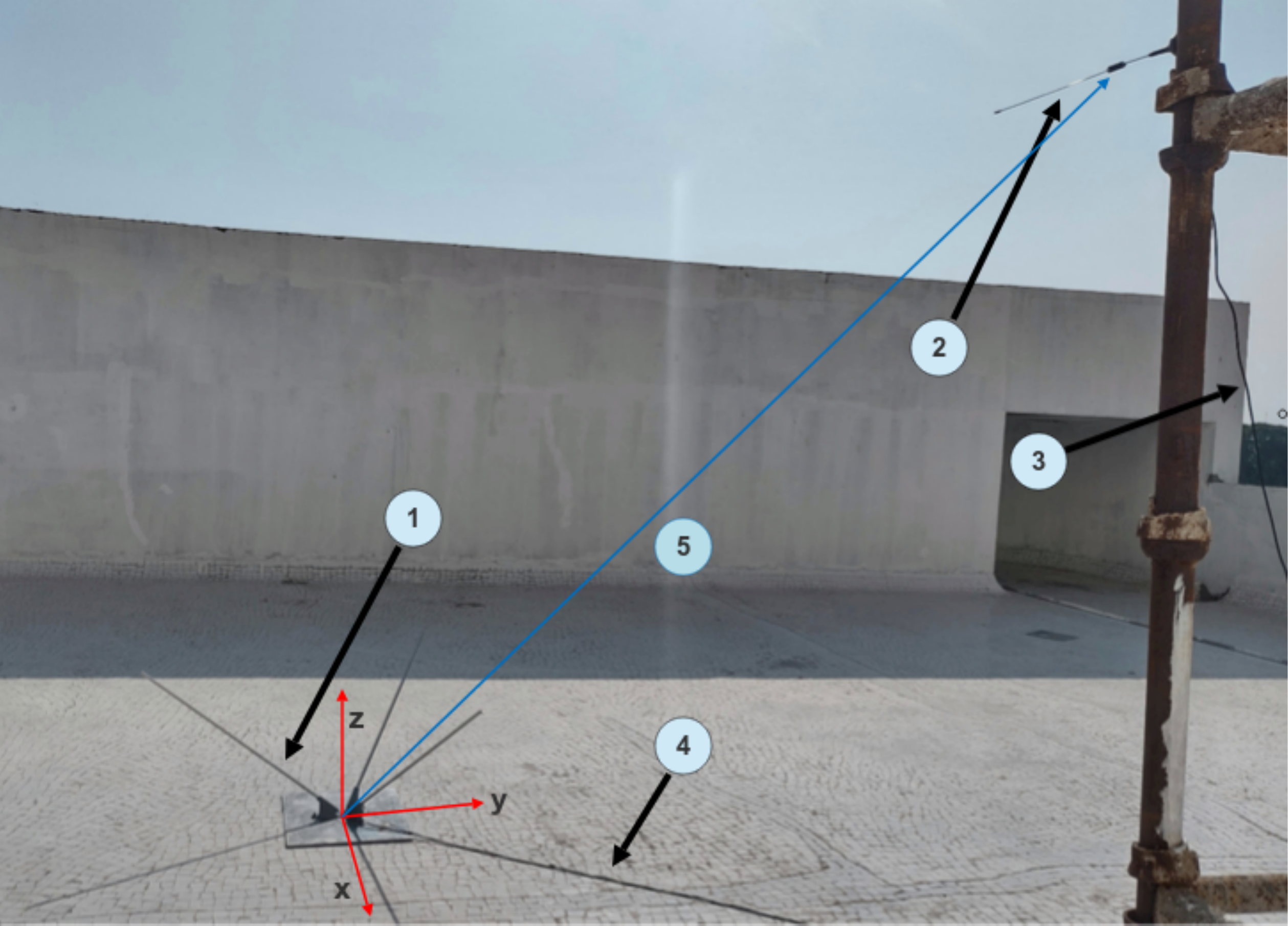}
    \caption{An image describing the experimental arrangement where, (1) is the Fabricated tripole antenna, the red arrows denotes the reference axis ($x$, $y$, $z$) and (2) is the synthetic source constitutes of a monopole and a RF generator\,(Keysight N5173B) set at frequency of 72\,MHz and power of 10\,dBm. (3) shows the RF line which connects monopole and RF generator. The data from the tripole at (1) is carried by RF line shown by (4) and is received using a DSO\,(Keysight Infinium DSO9254A). (5) indicates line of sight of the source which is at located at Az/El - $\sim$41$^\circ$/$\sim$51$^\circ$ from the reference axis displayed in the image.}
    \label{fig:Fig9}
\end{figure}

To test the algorithm, a proof of concept scaled version of DoA experiment was carried out at the resonant frequency of the antenna\,(length $\ge$ 1\,m).
This setup consists of a prototype of the tripole antenna arrangement, fabricated using a $\sim$1\,m long monopoles with resonance at $\sim$72\,MHz. 
The experiment is performed using a fabricated tripole antenna as receiving element\,($T_{RX}$) and a synthetic radio source made of an aerial antenna\,(Nooelec monopole antenna) that transmits a monotone signal\,(at 72\,MHz) generated by an RF generator from a known direction\,(Azimuth\,(Az) and Elevation\,(El)).
Figure \ref{fig:Fig7} shows the block diagram of the experimental setup.
The fabricated antenna is shown in Figure \ref{fig:Fig8}.
Figure \ref{fig:Fig9} shows the arrangement of the scaled experiment.
In this setup, a synthetic source is formed using a Keysight RF generator N5173B producing a monotone signal at 72\,MHz with +10\,dBm power, connected to the aerial antenna.
The receiving system consists of a tripole antenna connected directly to a digital storage oscilloscope\,(DSO) via RF cable to keep the phase distortions due to inline components (like amplifiers and filters) to a minimum.

\begin{figure}[!ht]
    \centering
    \includegraphics[width=1\linewidth]{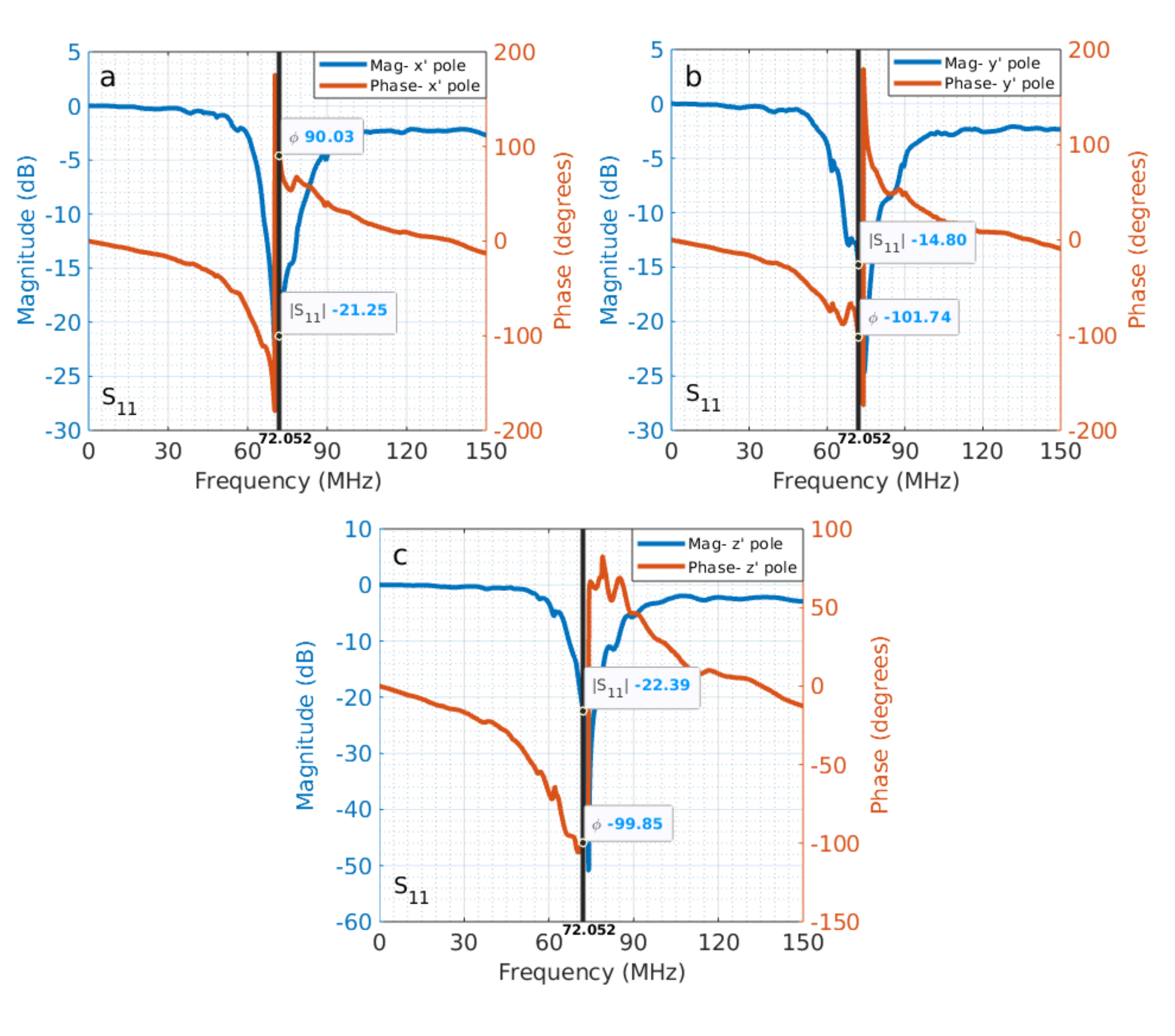}
    \caption{Complex Reflection coefficient measurement of all the monopole elements in the tripole antenna wherein, the black vertical marker shows the resonance frequency of 72 MHz.}
    \label{fig:Fig10}
\end{figure}

\begin{figure}[h]
    \centering
    \includegraphics[width=1\linewidth]{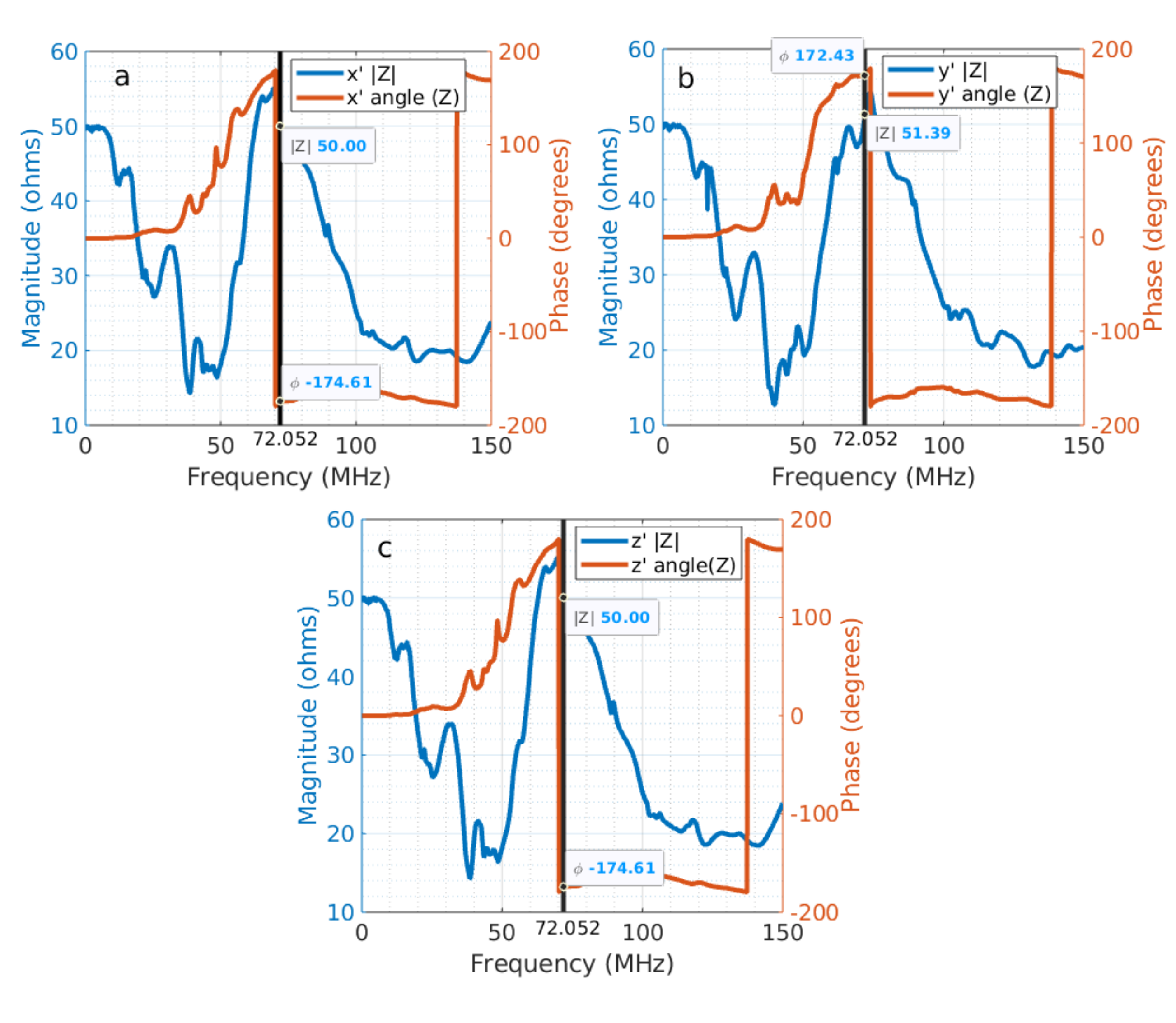}
    \caption{Radiation Impedance of all the elements in the tripole antenna along the local reference axis $x'$, $y'$, and $z'$ as illustrated in Fig. \ref{fig:Fig8}.}
    \label{fig:Fig11}
\end{figure}

Figure \ref{fig:Fig10} and \ref{fig:Fig11} show the tripole antenna's reflection coefficient\,($S_{11}$) and Radiation Impedance of each element in the configuration demonstrating its resonance at $\sim$72\,MHz.
This experiment was carried out utilizing minimum circuit components, since addition of circuit component would contaminate the phase of the received signal (see Appendix \ref{secA1}).

The dataset of this experiment has been recorded with the sampling a frequency of 8\,GHz in the DSO.
The dataset is then down sampled by a factor of 4 to improve the frequency resolution\,($F_r$) of the data.
The resultant $F_s$ = 2\,GHz corresponds to $F_r$ = 0.98\,MHz for 2048 point FFT.

\begin{figure}[h]
    \centering
    \includegraphics[width=1\linewidth]{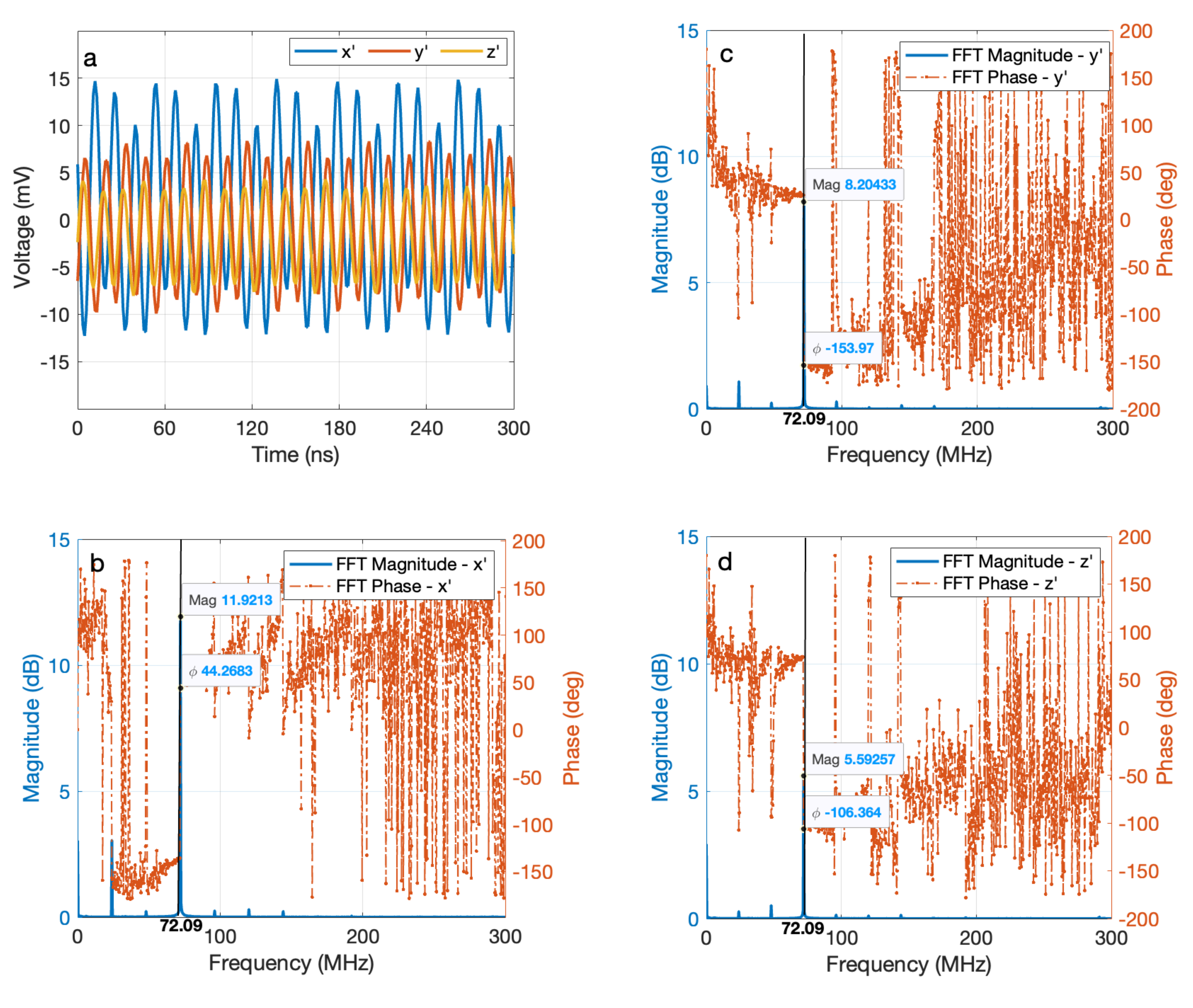}
    \caption{Signal received from a synthetic source of $\sim$72\,MHz having Azimuth of $\sim197^\circ$ and elevation of $\sim24^\circ$. Here, (a) shows high time resolution voltage values received from the source using the Tripole, (b) to (d) shows the Amplitude and phase of the received signal in frequency domain by the tripole antenna along the local reference axis $x'$, $y'$, and $z'$ as illustrated in Fig. \ref{fig:Fig8}.}
    \label{fig:Fig12}
\end{figure}

\begin{figure}[h]
    \centering
    \includegraphics[width=1\linewidth]{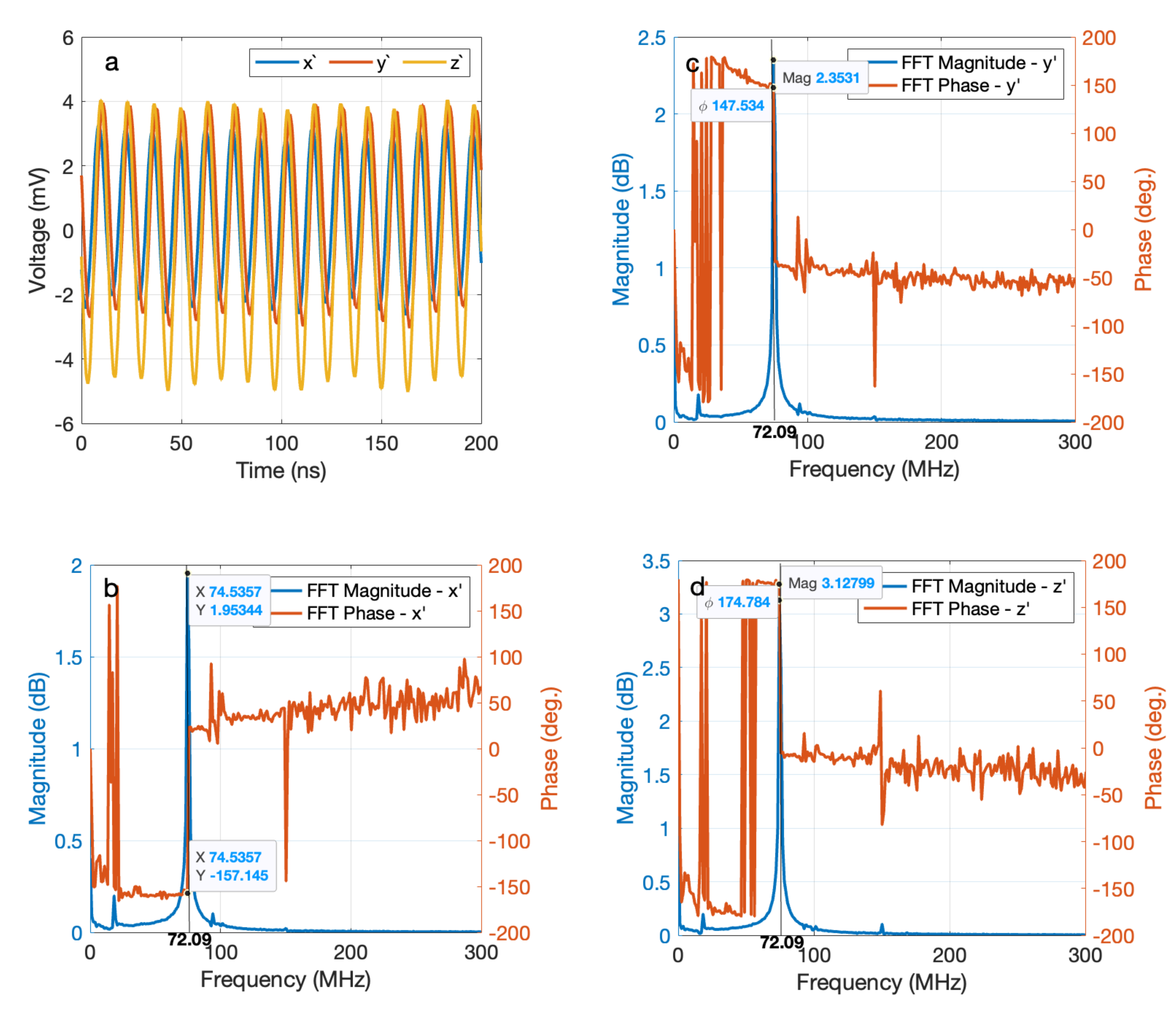}
    \caption{Signal received from a synthetic source of $\sim$72\,MHz having Azimuth of $\sim210^\circ$ and elevation of $\sim29^\circ$.}
    \label{fig:Fig13}
\end{figure}

\begin{figure}[!ht]
    \centering
    \includegraphics[width=1\linewidth]{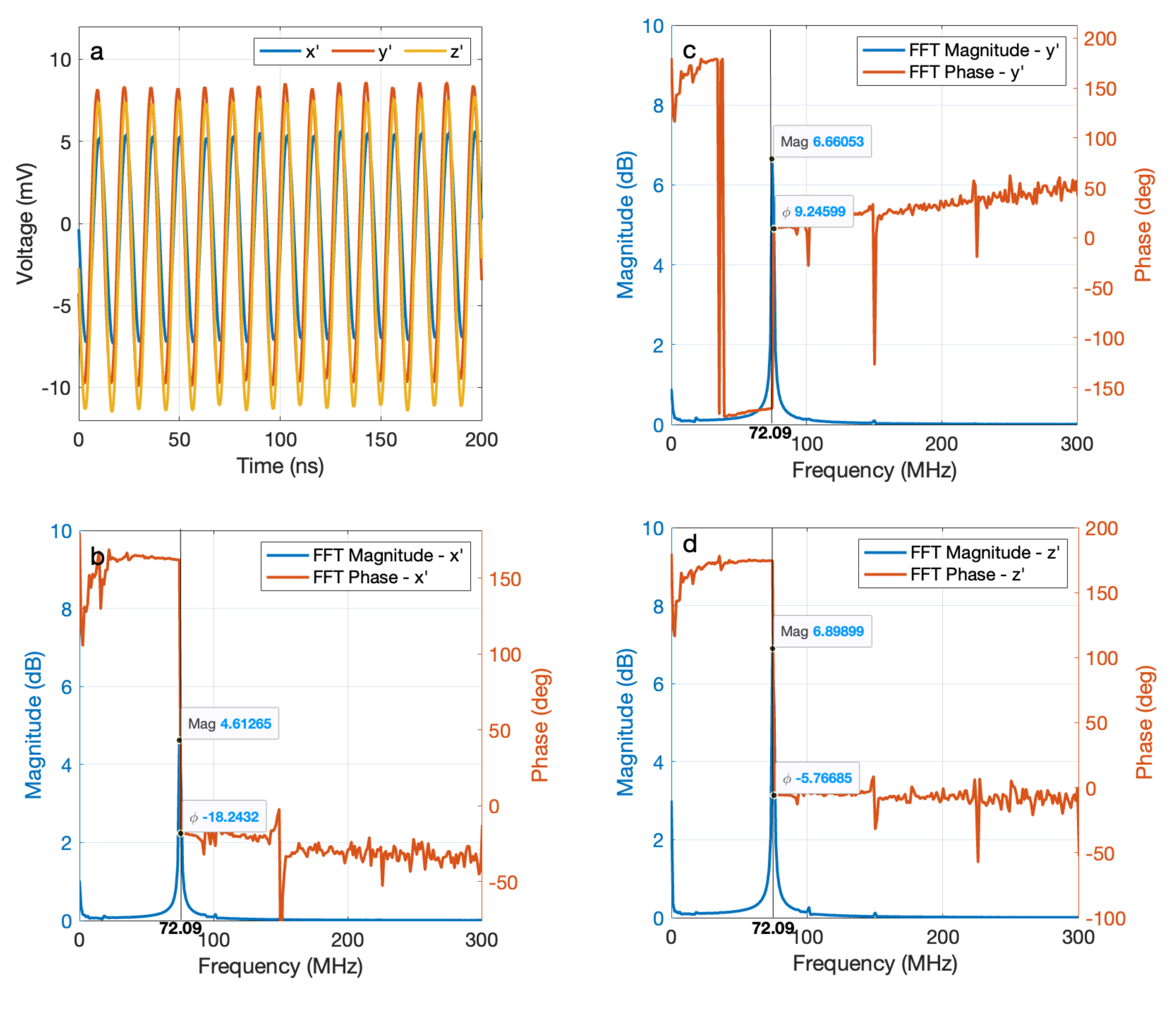}
    \caption{Signal received from a synthetic source of $\sim$72\,MHz having Azimuth of $\sim41^\circ$ and elevation of $\sim51^\circ$.}
    \label{fig:Fig14}
\end{figure}

Figure \ref{fig:Fig12}, \ref{fig:Fig13}, and \ref{fig:Fig14} shows the received signal in time domain and frequency domain for the synthetic source located at Az/El of $\sim197^\circ$/$\sim24^\circ$, $210^\circ$/$29^\circ$, and $\sim41^\circ$/$\sim51^\circ$. 
It is evident from the time and frequency domain plots in figures \ref{fig:Fig12} to \ref{fig:Fig14} that the amplitudes and the phases received by the three elements in the antenna are different\footnote{The url - ``https://drive.google.com/file/d/1lldWOl5q1jK\_3br4wBW0qT0y9olMdE8q/view?usp=sharing" contains a video showing the phase shift in the received signal if the source is mobile.}.
It is observed that there is a abrupt jump in the received signal phase. 
This phase jump maybe related to the received amplitude of the signal.
For example, in Figure \ref{fig:Fig12}(a), the $x'$ and $z'$ antennas have the highest and the smallest amplitude.
Correspondingly the highest and lowest step jump are seen in Fig. \ref{fig:Fig13}(b) and Fig. \ref{fig:Fig13}(d) respectively.
The stated phenomenon can be observed in readings from all three directions in this laboratory test.
Based on the discussion in section \ref{theory} and \ref{Results}, it is shown that the difference in amplitude and phases between the antennas is utilized for estimating the DoA.
From figures \ref{fig:Fig12} to \ref{fig:Fig14} it is observed that the change in the direction of the sources causes a change in amplitude and phases of the received signal.
In the estimate algorithm, the phase distortion caused by the antenna and the transmission cable impedance has to be considered\,(see Appendix \ref{secA1}). 
Oscilloscope calibration with the transmission cable has been performed to account for reading errors by oscilloscope and the phase error due to transmission cable.
This removed the losses due to the cable impedance as it has been calibrated along with the oscilloscope.
Thus, the modification in the algorithm was done to consider the phase distortion by the antenna impedance.

\begin{table}[h!]
    \centering
    \begin{tabular}{p{4cm}|c|c}
        \hline
        Algorithm & Estimated Az/EL & Actual Az/El  \\
        \hline
        Pseudovector estimation based DoA method \citep{Carozzi2000} & $212^\circ$/$15^\circ$ & \\
        MPM DoA method \citep{Sarkar1995, Yilmazer2006, Daldorff2009, Chen2010} & $202^\circ$/$18^\circ$ & $\sim197^\circ$/$\sim24^\circ$ \\
        SAM DoA method & $202^\circ$/$18^\circ$ &  \\
        \hline
        Pseudovector estimation based DoA method \citep{Carozzi2000} & $207^\circ$/$32^\circ$ &  \\
        MPM DoA method \citep{Sarkar1995,Yilmazer2006,Daldorff2009,Chen2010} & $213^\circ$/$31^\circ$ & $\sim210^\circ$/$\sim29^\circ$ \\
        SAM DoA method & $212^\circ$/$31^\circ$ &  \\
        \hline
        Pseudovector estimation based DoA method \citep{Carozzi2000} & $43^\circ$/$48^\circ$ &  \\
        MPM DoA method \citep{Sarkar1995, Yilmazer2006, Daldorff2009, Chen2010} & $40^\circ$/$52^\circ$ & $\sim41^\circ$/$\sim51^\circ$ \\
        SAM DoA method & $40^\circ$/$52^\circ$ &  \\
        \hline
        
    \end{tabular}
    \caption{A comparison of the DoA obtained from the Experiment using different methods applicable to SEAMS antenna configuration. The test has been carried out for three different radiation directions at resonant frequency. The ``Estimated Az/El" coloumn contains the estimated DoA by the algorithms and the ``Actual Az/El" column shows the physically measured DoA of the source. The emitted signal was of LP and has been detected in all the algorithm as LP; the SNR for this experiment was $>60$\,dB.}
    \label{tab:tab_exp}
\end{table}

Table \ref{tab:tab_exp} shows the estimated DoA for two different sources radiating at 72\,MHz with LP. 
The table also shows a comparative estimation between Pseudovector estimation based DoA method \citep{Carozzi2000}, MPM DoA method \citep{Sarkar1995, Yilmazer2006, Daldorff2009, Chen2010}, and SAM DoA method. 
In the experiment the signal is transmitted at the resonant frequency with a power of +10\,dBm having SNR of $>60$\,dB; it is observed that the error in the estimation is between 0\,-\,6$^\circ$ which is high for the given SNR.
However, it may be noted that all the DoA methods were able to characterise the EM wave as LP.
The large estimation error might be due to several factors affecting the experiment such as multi-path, non planar wave front due to source being close to the receiving antenna and the RFI environment of the laboratory. 
A more detailed study is required in order to understand the effect of physical or environmental parameters on the DoA estimation and hardware.
Similar efforts has been made to understand the arrival of waves from Saturn Kilometric Radiation by \cite{2006cecconi_PREVI}.
It is planned to carry out an elaborate experiment at our desired frequency band.
As discussed in section \ref{sim_set}, the SEAMS antenna will be an active antenna.
Thus, phase contamination in the received signal by the matching network of the antenna has to be adjusted as per its frequency response\,(described in Appendix \ref{secA1}).

\section{Conclusion}\label{Conclusion}

An optimized MPM DoA estimation algorithm by the addition of an averaging and polarization detection method\,(SAM DoA) is described in this paper.
The salient features of the SAM DoA are shown in table \ref{tab:tab1}.
\begin{table}[!ht]
    \centering
    \begin{tabular}{c|c|c}
        \hline
        Features & MPM DoA  & SAM DoA  \\
        \hline
        RMSE in az/ele @ SNR 10dB & 0.998/0.97 & 0.35/0.77 (for \textit{n}=20)\\
        Polarization detection & Yes & Yes \\
        RHCP or RHEP detection & Yes & Yes \\
        LHCP or LHEP detection & No & Yes \\
        Preferred Antenna configuration & Tri-dipole & Tripole\\
        \hline
        
    \end{tabular}
    \caption{A comparison table of Matrix Pencil Method\,(MPM) DoA algorithm and the proposed SAM DoA Algorithm}
    \label{tab:tab1}
\end{table}

The averaging algorithm estimates the mean of multiple FFT snapshots before applying the MPM to reduce noise and improves the estimation of the incoherent frequencies in the given spectra.
The Polarization detection method enabled the detection of different polarisations (Table \ref{tab:pol_dec}).
With these improvements, the algorithm has become adaptable so that it can respond to the increase in the number of incident incoherent waves.
In addition, this algorithm has multiple applications in remote sensing where the polarisation of the reflected wave is important, as in the case of agriculture, cracks in metarials, etc\,\citep{egido2012_RS, dvorsky2020_IEEE}.

The present analysis elaborates the simulation carried out with tri-dipole and tripole antenna configurations.
The simulation results, show that the signal received by both the configurations is the same considering the direction constraints of our simulation setup.
This will be used in the SEAMS mission which is proposed as an orbiter mission for the far-side of the moon and the antenna for such low frequencies will be an active antenna.

The total computational complexity of SAM DoA has increased when compared with the other algorithms\,(Table \ref{tab:tab3}).
However, the computational cost remains lower than that of other well known algorithms such as MUSIC, M-MUSIC, MP-MUSIC, ESPRIT and many more \citep{Gentilho2019}. 
The choice of modifications in the algorithm are done such that the computational cost remains low and most of the processing can be performed by the onboard computational devices or FPGA.
This is necessary due to the data transfer limitations in space \citep{Walker2013}.

A proof of concept scaled experiment of DoA carried out in our laboratory at the resonant frequency of the antenna validated the feasibility of detection of DoA by utilizing a triaxial antenna configuration, but in order to test the performance of different algorithms, extensive tests are required.
In the laboratory experiment, the observed phase variations before and after the received frequency\,(Fig. \ref{fig:Fig12} to \ref{fig:Fig14}) could be due to noisy environment or test equipments used.
In order to better understand the phase variations, the data needs to be recorded using a sensitive data logger's ADC dump and then analyzed.

\section*{Acknowledgements}
H.A.T. acknowledges the valuable discussions with Mr.\,Atharva Kulkarni (SPPU) regarding the SEAMS payload design and electronics and with Mr.\,Krishna Makhija (NRAO) regarding the CST simulations. Authors are thankful to Department of Electronic Science, SPPU (specially Prof D. Gharpure) for its support right from the beginning of this project\,(2017). Authors are thankful to the entire team of the SEAMS project. H.A.T. is thankful to Mr.\,Archisman Guha (IIT Indore) and Mr.\,Abhijeet Dutta (IIT Indore) for their support in DoA experiment. H.A.T. is thankful to research scholars Ms.\,Aishrila Majumder (IIT Indore), Ms.\,Deepthi Ayyagari (IIT Indore) and Mr.\,Sarvesh Mangla (IIT Indore) for their technical suggestions while drafting this manuscript. Authors also thank Dr.\,C. Bhatacharya for his critical comments. 

\begin{appendices}

\section{Phase Contamination by electronic components}\label{secA1}

\begin{figure}[!ht]
    \centering
    \includegraphics[width=0.6\linewidth]{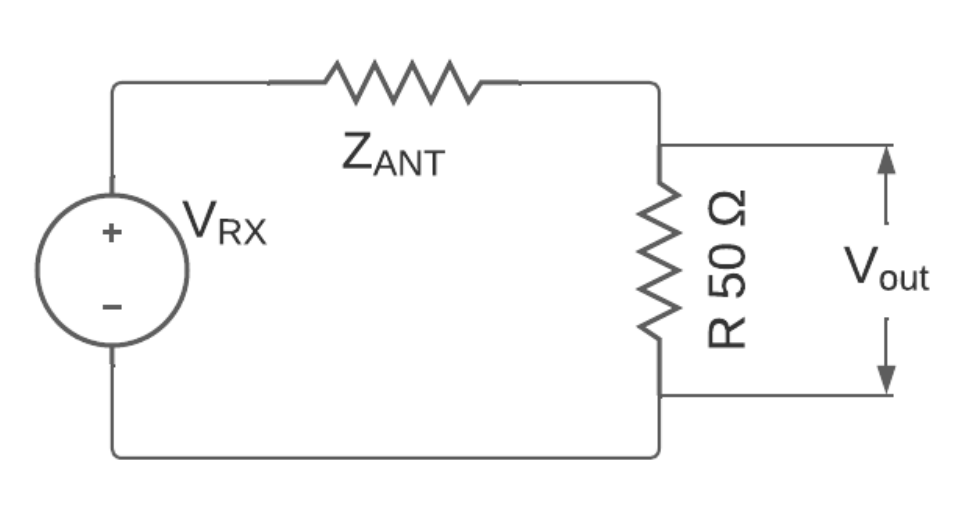}
    \caption{Receiving antenna circuit equivalent. $V_{RX}$ is the voltage received by the antenna, $Z_{ANT}$ is the intrinsic impedance or radiation resistance of the antenna, $R$ is the load resistance of 50\,$\Omega$, and $V_{out}$ voltage across the load.}
    \label{fig:Fig15}
\end{figure}

In circuit theory any receiving antenna can be viewed as a independent voltage source with a source impedance called antenna impedance or radiation resistance \citep{Balanis-2012-antenna}.
Figure \ref{fig:Fig15} is the circuit equivalent diagram of an receiving antenna with a load resistance of 50\,$\Omega$.

Since, the voltage received\,($V_{RX}$) by the antenna is due to electric field of EM wave then, $V_{RX}\,=\,h_{eff}\vec{E}$ where $h_{eff}$ is the effective height of the antenna and $\vec{E}$ is the electric field present in the EM wave. 
Using the plane wave consideration the electric field component can be written as $\vec{E}\,=\,E_0 e^{j(\vec{k}\cdot\vec{r} - \omega t)}$ here, $\omega\,=\,2\pi f$.
Thus, the voltage received can be written as following:
\begin{equation}\label{eq:a1}
    V_{RX} = h_{eff}E_0 e^{j\vec{k}\cdot\vec{r}} e^{-j \omega t} = A e^{-j \omega t}
\end{equation}

Using equation \ref{eq:a1} and circuit in Figure \ref{fig:Fig15} the received signal $V_{out}$ can be written as 
\begin{equation}\label{eq:a2}
    V_{out} = \frac{V_{RX}\times50}{50+Z_{ANT}}
\end{equation}

As impedance comprises of resistive\,($R$) and reactive component thus antenna impedance can be written as $Z_{ANT}\,=\,R_{ANT}+jX_{ANT}$.
Considering the antenna impedance and equation \ref{eq:a2} on can observe analytically how phase is being modified due to the impedance in equation \ref{eq:a3}.
\begin{equation}\label{eq:a3}
    V_{out} = \frac{50 A}{\sqrt{R_{ANT}^2+50^2}} e^{-j[wt + tan^{-1}({X_{ANT}/(R_{ANT}+50)})]}
\end{equation}

In case of addition of several circuit components either in series or in parallel, the antenna impedance $Z_{ANT}$ in equation \ref{eq:a3} has to be replaced by the effective impedance of the circuit also known as the Thevenin's equivalent.

\section{Matrix Pencil Method}\label{secA2}
The Matrix Pencil method is used to obtain the best estimates since it interacts directly with the data instead of generating a co-variance matrix, reducing computer complexity \citep{Yilmazer2006}. 
Eq. (\ref{eqn:eqn6}) is used to generate a Hankel matrix in order to estimate N and $\omega^n$. 
\begin{equation*}
\begin{aligned}
\scalebox{0.8}{$\Lambda$} =
\scalebox{0.75}{
$\begin{bmatrix}
S(0) & S(1) & \cdots & S(L)\\
S(1) & S(2) & \cdots & S(L+1)\\
\vdots & \vdots & \ddots & \vdots\\
S(M-L-1) & S(M-L) & \cdots & S(M-1)
\end{bmatrix}_{(M-L) \times (L+1)} $}
\end{aligned}
\end{equation*}
where, $L$ is selected between $(M/3,M/2]$ for optimum performance and is known as the pencil parameter \citep{Sarkar1995}; $M$ is the total sample length.
The real matrix($\Lambda_R$) is computed using a Unitary matrix transformation \citep{Sarkar1995}\,({ \small $\Lambda_R = U^\dagger [\Lambda \mid \Pi_{M-L} \Lambda^* \Pi_{L+1}]U$; where, $^\dagger$ represents hermitian conjugate and $U$ is the unitary matrix \citep{Daldorff2009}}) and the complex number matrix $\Lambda$.
Later, an  estimate of the singular values of $\Lambda_R$ is generated using SVD formulation. 
Matrix $A_s$ consisting of $N$ largest singular vectors of $\Lambda_R$ is estimated by performing a thresholding operation on the normalized Eigen value i.e., $\sigma_i/\sigma_{max}$.
$N$ generalized singular values are then calculated using an unitary transformation\,($-[Re(U^\dagger J_1 U)A_s]^{-1}\cdot Im(U^\dagger J_1 U)A_s$) which are, $\psi_1, \psi_2,\cdots,\psi_N$.
Also, $N$ incoherent frequencies are calculated by $\omega^n = 2arctan(\psi_n)/\delta$ for $n = 1,2,\cdots,N$ \citep{Daldorff2009,Chen2010,Harsha2021ieee}.

\end{appendices}

\bibliographystyle{ACM-Reference-Format}
\bibliography{main.bib}  






\end{document}